\documentclass[12pt]{article}
\usepackage[a4paper,bindingoffset=0.in,%
left=0.6in,right=0.6in,top=1.in,bottom=0.8in,%
footskip=.3in]{geometry}
\usepackage{blindtext}
 
\usepackage[scaled]{helvet}
 
\usepackage[T1]{fontenc}

\usepackage[utf8]{inputenc}
\usepackage{authblk}
\usepackage[]{lipsum}
\usepackage{fancyhdr}
\usepackage{lmodern}
\usepackage[utf8]{inputenc}
\usepackage[english,activeacute]{babel}
\usepackage{mathtools}
\usepackage{amsmath,amssymb,amsthm}
\usepackage{graphicx}
\usepackage{enumerate}
\usepackage{xypic}
\usepackage{hyperref}
\usepackage{stmaryrd}
\usepackage{wrapfig}
\usepackage[font=small,hang]{caption}
\usepackage{enumitem}
\setlist[enumerate,1]{start=1}
\usepackage{tikz,kantlipsum}
\usetikzlibrary{tikzmark}
\usepackage{color}
\usepackage{afterpage}

\usepackage{url,hyperref,lineno,microtype,subcaption}
\usepackage[onehalfspacing]{setspace}

\usepackage{mathscinet}

\usepackage[backend    = biber,
            sortcites  = true,
            giveninits = true,
            doi        = false,
            isbn       = false,
            url        = false,
            sortlocale = auto,
            nohashothers = false,
            maxnames = 100,
            style      = numeric]{biblatex}

\usepackage{fancyhdr}
\makeatletter
\let\ps@plain\ps@fancy
\makeatother
\renewcommand{\footnoterule}{%
  \kern -5pt
  \hrule width \textwidth height 0.1pt
  \kern 5pt
}

\usepackage{url,hyperref,lineno,microtype,subcaption}
\usepackage[onehalfspacing]{setspace}
\makeatletter
\renewcommand{\maketitle}{\bgroup\setlength{\parindent}{0pt}
\begin{flushleft}
  \textbf{\@title}
  \@author
\end{flushleft}\egroup
}
\makeatother

\usepackage[format=plain]{caption}

\title{\LARGE A Mesoscopic Lattice Model for Morphology Formation in Ternary Mixtures with Evaporation}
\date{}
\author{\\ \vspace{0.5cm}
\textbf{Mario Setta}\,$^{1,2}$, \textbf{Vì C. E. Kronberg}\,$^{3}$, \textbf{Stela Andrea Muntean}\,$^{4}$, \textbf{Ellen Moons}\,$^{4}$, \textbf{Jan van Stam}\,$^{5}$, \textbf{Emilio N. M. Cirillo}\,$^{6}$, \textbf{Matteo Colangeli}\,$^{1}$, and \href{mailto:adrian.muntean@kau.se}{\textbf{Adrian Muntean}}\,$^{\star,2}$ \\ \vspace{0.5cm}
$^{1}$\textit{Dipartimento di Ingegneria e Scienze dell'Informazione e Matematica, Università
degli Studi dell'Aquila, L'Aquila, Italy} \\
$^{2}$\textit{Department of Mathematics and Computer Science, Karlstad University, Karlstad,
Sweden}\\
$^{3}$\textit{Department of Mathematics and Computer Science, Eindhoven University of Technology, Eindhoven, The Netherlands}\\
$^{4}$\textit{Department of Engineering and Physics, Karlstad University, Karlstad,
Sweden}\\
$^{5}$\textit{Department of Engineering and Chemical Sciences, Karlstad University, Karlstad,
Sweden}\\
$^{6}$\textit{Dipartimento di Scienze di Base e Applicate per l'Ingegneria, Facoltà di Ingegneria Civile e Industriale, Sapienza Università di Roma, Italy} \\\vspace{0.5cm}
Correspondence$^{\star}$: \\
Adrian Muntean \\
\underline{\href{mailto:adrian.muntean@kau.se}{adrian.muntean@kau.se}}
}

\begin{document}

\pagenumbering{arabic}

\maketitle
\vspace{4cm}
\noindent
\begin{abstract}
We develop a mesoscopic lattice model to study the morphology formation in interacting ternary mixtures with evaporation of one component. As concrete application of our model, we wish to capture morphologies as they are typically arising during fabrication of organic solar cells. In this context, we consider an evaporating solvent into which two other components are dissolved, as a model for a 2-component coating solution that is drying on a substrate. We propose a 3-spins dynamics  to describe the evolution of the three interacting species. As main tool, we use a Monte Carlo Metropolis-based algorithm, with the possibility of varying the system's temperature, mixture composition, interaction strengths, and evaporation kinetics. The main novelty is the structure of the mesoscopic model -- a bi-dimensional lattice with periodic boundary conditions and divided in square cells to encode a mesoscopic range interaction among the units. 
We investigate the effect of the model parameters on the structure of the resulting morphologies. Finally, we compare the results obtained  with the mesoscopic model with corresponding ones based on an analogous  lattice model with a short range interaction among the units, i.e. when the mesoscopic length scale coincides with the microscopic length scale of the lattice.

\vspace{1cm}
\end{abstract}

\newpage

\section{Introduction}
The mechanisms underlying mesoscopic and macroscopic pattern formation from local microscopic interactions are explored in many fields of physics, chemistry, and biology \cite{Global}. Lattice–based modelling of interactions between units\footnote{In the next sections, we will refer to these units as ``particles''. Note that "particles" do not necessarily mean physical entities like atoms or molecules. They should rather be regarded as interaction sites at the microscopic level.} (magnetic spins, agents, molecules, pedestrians, colloids, etc.) can give a coherent description of real behavior in many different situations. Well-known examples include  descriptions of phase transitions, flame propagation, spinodal decomposition, formation of magnetization bands, acceleration shock waves in traffic flow, building of coherent groups in large pedestrian crowds and molecules moving inside a cell searching for exit gates; see e.g. \cite{Complex,complexity} and references cited therein. In this paper, we focus on a setting inspired from work done on organic solar cells (compare e.g. \cite{Jan_Ellen}), where the effect of the evaporation of a solvent -- background environment for a mixture of two interacting polymers -- on the formation of polymer-polymer stable mesoscopic configurations, called here {\em morphologies}, is of strong interest.  From the energy harvesting perspective, this is a particularly relevant subject, since one expects that the shape and spatial arrangement of morphologies can affect considerably the overall power conversion efficiency of organic solar cells \cite{Jeroen, CondMat, YeLo}.  Conceptually related situations arise in the dynamics of interacting populations driven by different opinions and targets \cite{Saintier}. According to \cite{complexity}, both these applications belong to the realm of ``complexity''. %

Using the framework offered by  lattice-based modeling (see e.g. \cite{Ritter,MCMC} for the general methodology and \cite{Friedli} for a review of the theoretical foundations), we aim at understanding to which extent the formation of stable spatial patterns (morphologies), which are obtained as a result of pair-wise interactions in a ternary mixture with one evaporating component, depend  on specific length scales higher than the grid size of the lattice.  We refer to such larger scales as mesoscales and we label them by $\lambda$.  The interest in unveiling mesoscale effects  was triggered by our previous simulation  results reported  in \cite{Cirillo,Simon} and  \cite{SAM}, where we noticed the occurrence of  different types of morphology shapes. The simple observation that the geometry of the shapes depends on the choice of model parameters (the system's temperature, volatility, interaction parameters, etc.) makes us wonder whether our simulation  results are going to be drastically different if the overall dynamics combines information not only from a microscopic scale, but from both microscopic ($\lambda=1$) and mesoscopic length scales ($1<\lambda\ll L,$ with $L$ denoting some macroscopic length scale, e.g. the size of the simulation box). In particular, the vast separation of micro-, meso- and macro-scales is often invoked in the mathematical derivations of macroscopic behavior, used in statistical mechanics and in kinetic theory of gases as described for instance in \cite{limits,cerci,col09}.

Note that in \cite{Cirillo} we explored the parameter space leading to morphologies as obtained with simulations done exclusively at the microscopic scale (i.e., pair-wise interactions involving only the nearest neighbours of randomly selected lattice sites). Further motivation towards performing mesoscale-level simulations in a closely related context is mentioned in \cite{Jasper_mesoscale}. %

In this paper, we propose a two-scale lattice model capable of producing morphologies, where the interactions within the mixture capture not only microscopic information (from pairs of spins) but also mesoscopic information (from pairs of $\lambda$-sized blocks of spins). The structure of such a two-scale model is inspired by the setting considered in Ch. 4 of  \cite{limits}, where one considers systems characterized by having the inter-atomic and mesoscopic characteristic interaction lengths sharply separated.  In this context, the discussion is done in terms of Kac potentials; compare {\em loc. cit.}. Potentially, such an approach can provide an  alternative two-scale model for which the so-called Lebowitz-Penrose mean-field limit might be proven rigorously, at least in the absence of the evaporation process, which poses additional mathematical challenges as it is a non-equilibrium interface process; see \cite{LP66} for details on the Lebowitz-Penrose scaling of  Hamiltonians  and \cite{limits} for suitable mathematical techniques to study rigorously the passage to the continuum limit. We will study elsewhere the passage to the hydrodynamic limit in our setting. As we will see in Section \ref{model}, the proposed model incorporates both short-distance and long-distance interactions. Interestingly, a  non-intuitive effect stands out -- if a relatively low amount of  solvent (evaporating component) is present in the mixture, then our two-scale model exhibits morphologies with self-similar features, i.e. one is able to zoom in and zoom out inside the geometry of the morphologies by suitably varying the intermediate scale $\lambda$ together with a proportional modification of the size of the simulation box.
This effect is a direct consequence of the scaling choice of the Hamiltonian functional driving the dynamics. On the other hand, with the current scaling we are unable to obtain mean-field effects. More work is needed in this direction, especially if one is tempted to find a rigorous link between the  morphologies obtained with microscopic and/or mesoscopic lattice models, as done here, with the ones obtained by means of coupled systems of Cahn-Hilliard-type models, as for instance  was performed in \cite{Wodo,Jasper_Charlie,Benoit}  and references cited therein.

The paper is organized as follows: In Section \ref{model}, we describe the proposed mesoscopic lattice model. 
The main work consists in performing simulation tests and interpreting the results. In Section \ref{basic}, we show basic simulation results obtained for a fixed value of $\lambda$, bringing the attention to a particular mesoscopic level. To fix ideas, we have selected $\lambda=4$. We are using this scenario to explore the effect of various parameters (mixture composition, temperature of the system, volatility, interaction strengths among mixture components) on the formation of morphologies. This type of numerical results show that our mesoscopic model is able to capture the type of results obtained in \cite{Cirillo}. Additionally, a few typical mesoscopic features (like dependence of the morphology widths on $\lambda$ and quicker stabilization of morphologies) can now be pointed out - such features arise at each allowed choice of $\lambda >1$ and set the foundation for what we refer to as near self-similarity in morphology arrangements.  The role of Section \ref{multiscale} is to show specific effects that can now be probed varying the mesoscale length $\lambda$. Here we discover ways to investigate the interplay between changes in the model parameters and size effects, as well as multiscale effects by observing the basic simulation output (check, e.g., Section \ref{basic})  for distinct levels of $\lambda$. In the context of this section, we also study the connection between changing $\lambda$ and varying the box size of the simulations.  Section \ref{outlook} concludes the paper with a discussion of our main findings as well as with an outlook on possible further research concerning this type of interacting mixtures and related matters. 

\section{Model description}\label{model}

\begin{figure}[h!]
\begin{center}
    \includegraphics[width=0.6\textwidth]{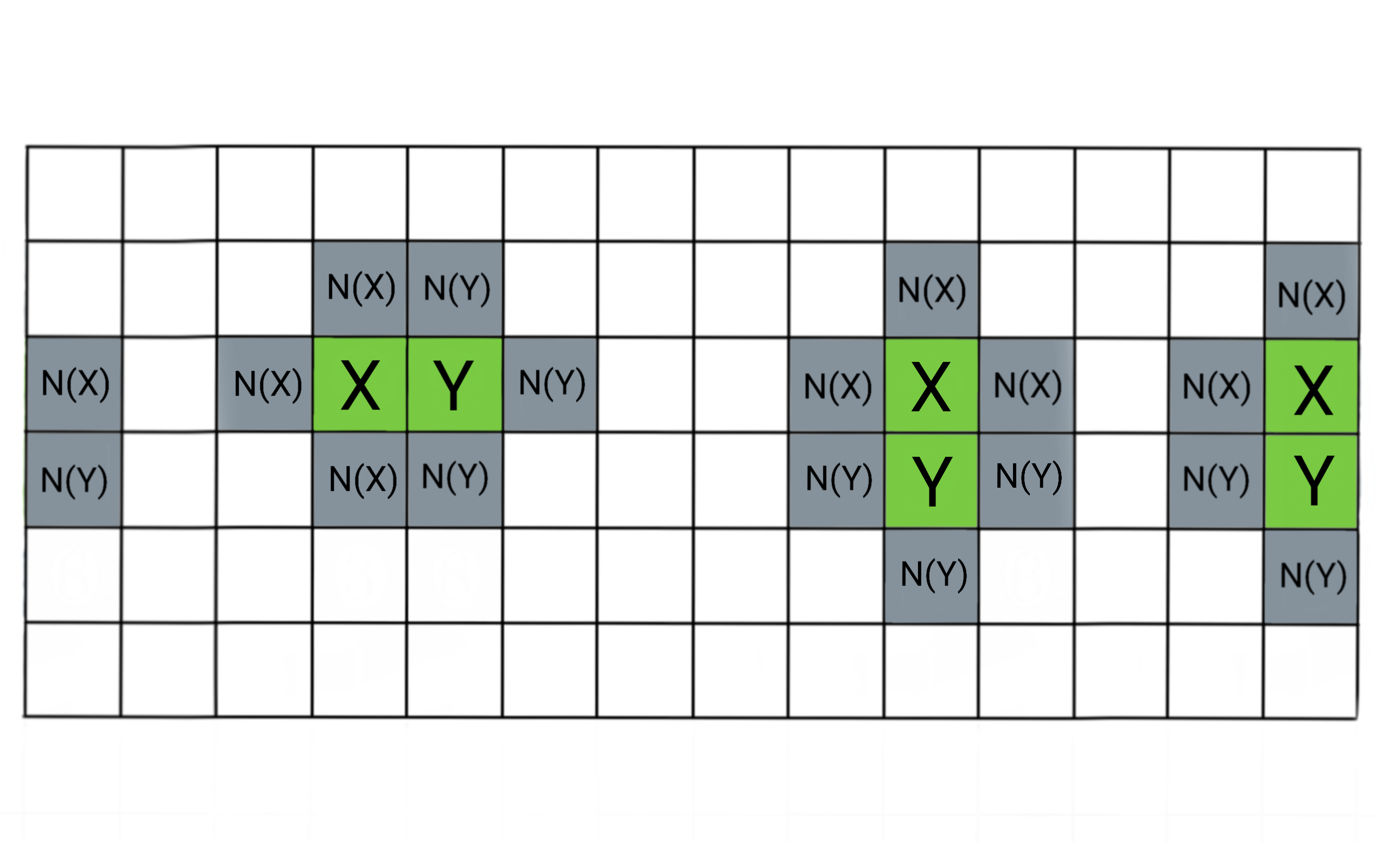}
    \caption{Example of cell partition of the lattice $\Lambda$ considering horizontal and vertical bonds, respectively. Periodic boundary conditions are exhibited. The cells in green are the ones involved in the spin-swap, namely the cells involved in the bond. The cells in a nearest neighbourhood of the bond are marked in grey. From the point of view of the Hamiltonian (\ref{ham_lambda}) we consider the bond $X,\ Y$ with nearest neighbours $\mathcal{N}(X) \smallsetminus \{Y\}$ and $\mathcal{N}(Y) \smallsetminus \{X\}$, respectively. Each group of colored cells is a subset $X \cup Y \subset \Lambda$.}
    \label{fig:lattice_neigh}  
\end{center}
\end{figure}

To build our mesoscopic lattice model, we consider a  two-dimensional rectangular lattice $\Lambda := \{ 1, \ldots, L_1 \} \times \{ 1, \ldots, L_2 \}$ with $L_1, L_2 \in \mathbb{N}$. We endow the lattice dynamics with periodic boundary conditions. An element of the torus $\Lambda$ is called \textit{site}. To reach mesoscopic descriptions, we introduce the long-distance interaction parameter $\lambda$. 
Concretely, we consider a partitioning of the lattice $\Lambda$ in square \textit{cells} of side $\lambda \ll \min\{L_1,L_2\}$. We end up with a lattice composed of $l_1 \times l_2$ squares containing $\lambda^2$ sites each, where $l_1 := L_1 / \lambda$ and $l_2 := L_2 / \lambda$. For computational convenience, we choose the values of $\lambda, L_1$, and  $L_2$ such that $l_1,l_2\in\mathbb{N}$. Any cell $X$ will be identified with the pair of integers $(X_1,X_2)\in\{ 1, \ldots, l_1 \} \times \{ 1, \ldots, l_2 \}$. Two cells $X,Y$ such that $X\not= Y$ are said to be \textit{nearest neighbors} if their euclidean distance is one. We refer to the set of the nearest neighbours of the cell $X$ as $\mathcal{N}(X)$. A pair of neighboring cells will be called \textit{bond}. Each site $x\in\Lambda$ has an associated \textit{spin variable} $\sigma_{x} \in \{ -1,0,+1\}$.
To model interaction we introduce the symmetric \textit{interaction tensor} $J\in\mathbb{R}^{3\times3}$ and we adopt the notation $J_{\alpha,\beta}$ with $\alpha,\beta=-1,0,+1$. We denote by $\sigma \in \{ -1,0,+1 \}^\Lambda$ any \textit{configuration} of the system on the lattice $\Lambda$ and by $\sigma_\Delta$ its restriction to $\Delta\subset\Lambda$. In Figure \ref{fig:lattice_neigh}, we show the subset $\Delta = X \cup Y$, composed by the cells $X,\ Y$ and their nearest neighbours, in different orientations. As main modelling step, given two neighboring cells $X$ and $Y$ and given two sites $x\in X$ and $y\in Y$, we consider the \textit{local energy} given by the \textit{Hamiltonian}
\begin{equation*}
    H_{x,y}: \sigma_{X \cup Y} \rightarrow \mathbb{R}^+
\end{equation*}
defined as
\begin{equation}
H_{x,y}(\sigma_{X \cup Y}) 
:= 
C \sum_{\genfrac{}{}{0pt}{2}{k \in X}{k \neq x}} 
J_{\sigma_{x}\sigma_{k}} 
+ C \sum_{\genfrac{}{}{0pt}{2}{k \in Y}{k \neq y}} 
J_{\sigma_{y}\sigma_{k}} 
+ 
\sum_{\genfrac{}{}{0pt}{2}{Z\in\mathcal{N}(X)}{Z\neq Y}} 
\sum_{\alpha, \beta} n_{\alpha}^{X} n_{\beta}^{Z} J_{\alpha \beta}
+ 
\sum_{\genfrac{}{}{0pt}{2}{Z\in\mathcal{N}(Y)}{Z\neq X}} 
\sum_{\alpha, \beta} n_{\alpha}^{Y} n_{\beta}^{Z} J_{\alpha \beta}
,
\label{ham_lambda}
\end{equation}
where for any $\alpha$ and $Z$ the notation $n_{\alpha}^{Z}$ refers to the number of particles of the species $\alpha$ in the cell $Z$, see Figure \ref{fig:lattice_neigh}.

In this case, the first two terms of the Hamiltonian capture 
the intra-cell energy, i.e. a mean field-like Hamiltonian restricted 
to the two cells $X$ and $Y$ (as in \cite{limits}) with respect to the fixed 
sites $x$ and $y$. 
This is given by the sum of all the interactions between the 
spins at $x$ and $y$ with all the other sites in the corresponding cells, 
according to the interaction tensor $J$. Hence, the first two terms in \eqref{ham_lambda} are the part of the Hamiltonian counting  
contributions from the microscopic scale. We involve a dimensionless tuning parameter $C$ to control the weight of the intra-cell energy 
with respect to the total energy $H_{x,y}$. 
The last two terms of the local Hamiltonian are the cell-cell 
interfacial energy, that represents the inter-cell energy. 
This part of the energy takes into account the interaction parameters of 
each combination of species weighted by the number of particles 
of those species in nearest neighbours cells and builds the mesoscopic contribution in the Hamiltonian. Since the Hamiltonian includes interactions at both microscopic and mesoscopic length scales, 
the resulting model is typically a multiscale model, or better 
said, a {\em micro-meso} model.

Usually, the interaction between particles of the same species is the one with minimum energy. Hence, to achieve the formation of  morphologies, we want the energy computed with this Hamiltonian to be decreasing during the evolution of the system.

We fix $p_{-1},\ p_0,\ p_{+1} \in [0,1)$ as initial probabilities to occupy one site such that $p_{-1} + p_0 + p_{+1} = 1$.
We choose the initial configuration\footnote{The number of iterations mimics the role of a discrete time variable $t \in [ 0, T_f ]$, where $T_f$ is the final "observation  time" corresponding to the last performed iteration.  Note that $t$ does not account for the real time. At every  $t$, the observed configuration is called $\sigma^t$. For convenience, the configuration at $t = 0$, i.e. $\sigma^0$, is called ``A'' in the code, while for $t > 0$, $\sigma^t$ is called $S$. Additionally, we define a configuration $S1$ for the code. The latter is the configuration $S$ with two swapped spins, as explained later on in this section. } $\sigma^0$ by setting any spin $\sigma^0_{x} = i,\ i \in \{-1,0,+1\}$ for all $ x \in \Lambda $  according to the probabilities $p_{-1},\ p_0$ and $p_{+1}$.
Sometimes, it will make sense to replace the probabilities $p_i$ with the actual percentages. %
We can now define $p_* := p_{+1}/(1-p_0)$. We make use also of two additional  parameters: the \textit{volatility}\footnote{The term ``volatility'' refers here to a drift leading the solvent particles towards the evaporation surface. It has nothing to do with the physical concept of volatility. The analogue name is though used since such break of symmetry applies only to one of the components of the mixture, i.e. to the potentially ``volatile'' one.}
parameter $\phi \in [0,1]$ and a parameter $\beta > 0$, such that $k_B T=\beta^{-1}$ is the \textit{thermal energy} of the considered system, involving the Boltzmann constant $k_B$.
Finally, we fix as \textit{stop parameter}, called $p_0^\star$, a certain percentage of the remaining solvent in the lattice. This stop parameter, called \texttt{stop} in the code, defines implicitly the final observation time $ T_{f}$ of our mesoscopic model. It is worth noting that in our approach we cannot link directly  the stop parameter with the correct estimation of the real time needed to build the morphology. Instead, to quantify the length of the simulations, we record the number of iterations needed to satisfy the stopping criterion. 
Therefore, the last configuration is denoted by $\sigma^{T_f}$ and will be the one with a percentage of remaining solvent less than or equal to the one defined by the stop parameter $p_0^\star$.

\begin{figure}[h!]
\begin{center}
    \includegraphics[width=0.8\textwidth]{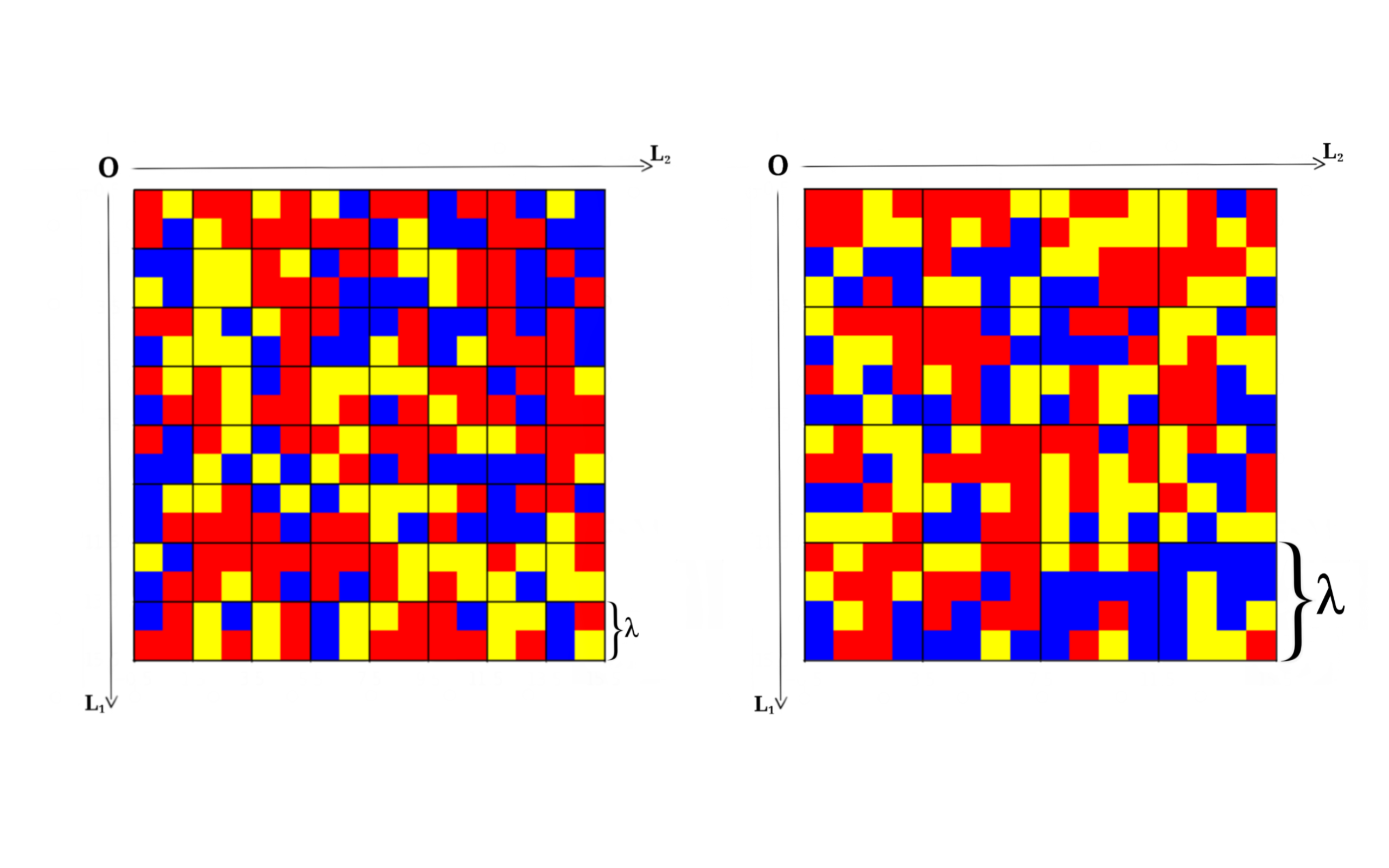}
    \caption{Orientation of the lattice $\Lambda$ for $L_1 = L_2 = 16$ and  cell sizes $\lambda = 2$ (left) and $\lambda = 4$ (right). The sites with yellow, red, and respectively, blue colors represent boxes containing particles with spin $-1$, $0$, and $+1$, respectively.} 
    \label{fig:lattice_l}  
\end{center}
\end{figure}
 We consider the lattice to be oriented as in Figure \ref{fig:lattice_l}. We divide the lattice in square cells of size  $\lambda \ll \operatorname{min}\{ L_1,\ L_2 \}$ to include the effects of a long distance interaction energy.
 The mathematical model is based on the \textit{discrete-time Markov chain Monte Carlo} method \cite{MCMC}. 
We propose a Metropolis-type algorithm, completed with volatility and evaporation rules, as follows:
\begin{enumerate}
    \item select a cell bond $(X,\ Y)$ in a uniformly random fashion;
    \item select a site for each cell in the bond uniformly at random, namely $x \in X$ and $y \in Y$;
    \item if $X = (l_1,\ X_2)$, $Y = (1,\ X_2)$ and $\sigma_y = 0$, then replace the zero with $+1$ with probability $p_*$, otherwise with $-1$. Update $p_*$ according to the new percentages of particles $p_{-1},\ p_0,\ p_{+1}$. This is the \textit{evaporation} of the zero (red) particle;
    \item if $X = (X_1,\ X_2)$, $Y = (X_1 + 1,\ X_2)$ and $y = 0$, with $X_1 < l_1$, then exchange the two spins with probability $\phi$. This is the \textit{volatility} of the zero (red) particle;
    \item otherwise swap the two spins, then consider the unswapped configuration $S$ and the swapped one $S1$ on $X \cup Y$. Compute $\Delta H = H_{x,y}(S1) - H_{x,y}(S)$ according to the Hamiltonian (\ref{ham_lambda}), that is the difference between the evolution of the Hamiltonian for the new and for the previous configuration. Accept the exchange with probability $1$ if $\Delta H < 0$ and with probability $\operatorname{exp}\{-\beta \Delta H / \lambda^2\}$ otherwise. This is the regular \textit{Metropolis} step.
\end{enumerate}
The process stops when the percentage of red particles (say $p_0$) is lower than the stop parameter defined above. We point out here an additional potentially important role of the parameter $\lambda$ that defines the size of the cells  (block spins, cf. the terminology from \cite{limits}) in the lattice, namely  $\lambda$ enters as tuning parameter in the probability $\operatorname{exp}\{-\beta \Delta H / \lambda^2\}$ computed in terms of $\beta$ and $\Delta H$. Essentially, we use the mesoscopic size of the cells to normalize the difference of energy with respect to the number of particles in a single cell, that is $\lambda^2$.

Note that a Markov chain is said to be \textit{ergodic} if it is possible to reach in a finite number of steps every configuration of the system starting off from an arbitrary initial configuration. Having this definition of ergodicity in mind, our system is not ergodic due to the the way we implement the evaporation part of the system. For our purposes, we assume that we can reach the configuration $\sigma^{T_f}$ in a finite number of steps, hence in a finite time interval.

The mesoscopic model proposed here is very much inspired by the one developed in  \cite{Simon} and is an extension of the lattice model proposed by Cirillo \textit{et al.} in  \cite{Cirillo}. In our earlier work, we referred to it as the $\lambda$-model. In the next sections, we explore by means of numerical simulations the capacity of our model to produce  morphologies when both short distance and large distance interactions interplay. The hope is to spot genuine mesoscopic effects by comparing our simulation output with previous work done in \cite{Cirillo, Simon, Mario} with microscopic versions of this model (i.e., when $\lambda=1$). 
It is worth mentioning already at this stage that our mesocopic model can be linked with what is observed based on purely microscopic descriptions, but is unable to reach mean-field information, i.e. when $\lambda= \min\{L_1,L_2\}$. %
One way to facilitate the latter connection, is to rescale suitably in terms of $\lambda$ the structure of the Hamiltonian, including changing the factor $C$ into $C(\lambda)$ as well as considering the simulation box as $\Lambda(\lambda)$. Inspiration from the Lebowitz-Penrose scaling indicated in \cite{limits} can be useful in our context. To understand this connection, a suitable asymptotic analysis needs to be performed. This is out of the scope of this paper.

\section{Basic simulation results}\label{basic}


In this section, we investigate the  effects produced by the different parameters on the shape and size of the formed morphologies.  The parameters involved in the simulations reported in this section are the ones independent of the interaction length scales in the system. Particularly, we analyze:
\begin{itemize}
    \item whether the phase separation is occurring or not, depending on the temperature of the system, by changing $\beta$;
    \item how the shape of formed morphologies  and the evaporation time are altered by the volatility parameter $\phi$ of the solvent;
    \item effects of the interaction parameter between the non evaporating species, i.e. $J_{+1,-1}$, on both the phase separation and the shape of morphology formations;
    \item how the initial ratio of species, given by the initial probabilities $p_{-1},\ p_0$ and $p_{+1}$, affects the shape of morphology formations.
\end{itemize}
To perform the simulations, we  build a code in Python  \cite{Langtangen,python}. We mainly use the module \href{https://numpy.org/doc/stable/}{NumPy}\footnote{\href{https://numpy.org/doc/stable/}{https://numpy.org/doc/stable/}} to build and modify the lattice and the module \href{https://matplotlib.org/contents.html}{Matplotlib}\footnote{\href{https://matplotlib.org/contents.html}{https://matplotlib.org/contents.html}} to print out the lattice as a figure.
In order to handle a large amount of simulations that require an intensive computational effort, we run our code on the supercomputer Kebnekaise, provided by High Performance Computing Center North (\href{https://www.hpc2n.umu.se/}{HPC2N}\footnote{\href{https://www.hpc2n.umu.se/}{https://www.hpc2n.umu.se/}}) of the Swedish National Infrastructure for Computing (\href{https://www.snic.se/}{SNIC}\footnote{\href{https://www.snic.se/}{https://www.snic.se/}}). 

Performing the simulations allows us to reach different morphology shapes at different iterations capturing the large time stationary behavior of our system. 
By  displaying graphically the obtained morphologies, we want to understand the role our parameters play in
the process of evaporation of the solvent through the surface of a thin film (described by our lattice), corresponding to the vertical cross section of a 3D box.
We choose the reference set of parameters to correspond to what is investigated in \cite{Mario} by means of a microscopic lattice-based model involving a short range interaction energy, that is $\lambda=1$. This allows us to test the code and facilitate eventual future comparisons of results.

\begin{figure}[h]
\centering
    \includegraphics[width=0.4\textwidth]{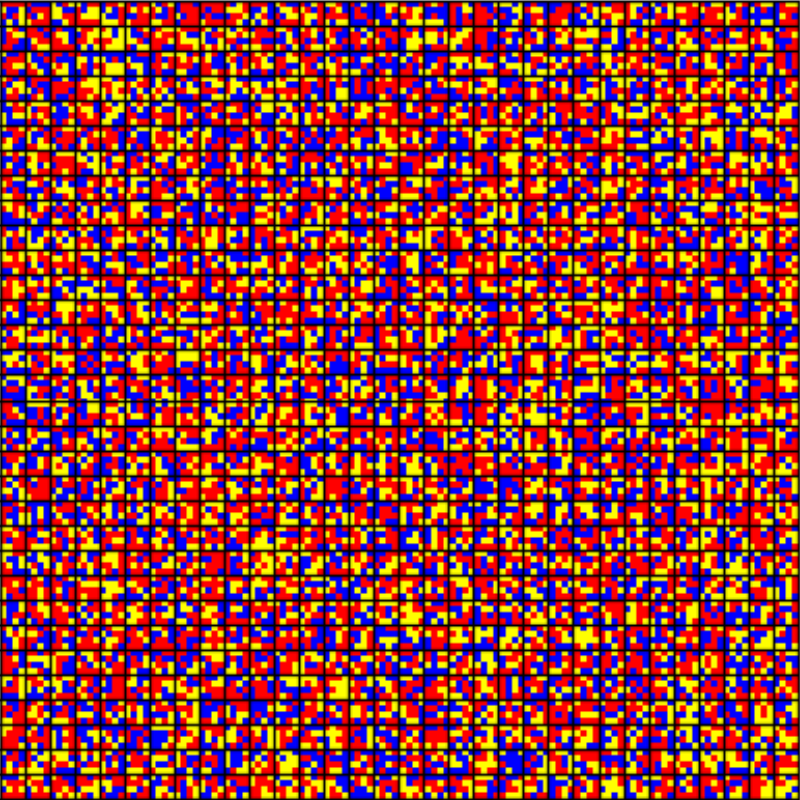}
    \caption{Initial configuration  $\sigma^0$ generated using initial proportion 40:30:30 for the red, blue and yellow particles, respectively. The considered lattice contains $128 \times 128$ particles, divided in square cells of size $\lambda = 4$. If not otherwise specified, each simulation starts from a configuration of this kind.}
    \label{fig:start}
\end{figure}

In this section, we consider a lattice containing $128 \times 128$ particles divided in yellow (with spin -1), red (with spin 0, the solvent that is evaporating), and blue (with spin +1), endowed with periodic boundary conditions. The size of the cells is fixed to be $\lambda = 4$.
The initial configurations are randomly generated according to the initial probabilities $p_0 = 0.4$, $p_{+1} = 0.3$ and $p_{-1} = 0.3$. This choice defines an initial mixture composition of 40:30:30 that is kept in the simulations.  In this case, the initial configurations are similar to the one shown in Figure \ref{fig:start}.
\vspace{0.6cm}
\newpage
The tuning parameter $C$ entering as factor in the structure of the  Hamiltonian (\ref{ham_lambda}) is fixed to be $C = 1$. Moreover, the energy parameters are set to be as indicated in the following symmetric tensor:
\begin{equation}
    J := 
    \begin{bmatrix}
    J_{-1,-1} & J_{-1,0} & J_{-1,+1} \\
    J_{0,-1} & J_{0,0} & J_{0,+1} \\
    J_{+1,-1} & J_{+1,0} & J_{+1,+1}
    \end{bmatrix} =
    \begin{bmatrix}
    0 & 1 & 6 \\
    1 & 0 & 1 \\
    6 & 1 & 0
    \end{bmatrix}.
    \label{interaction_simul}
\end{equation}
\begin{figure}[h!]
\centering
\noindent
\begin{tabular}{c c c c}
      $75\%$ solvent & $50\%$ solvent & $25\%$ solvent & $10\%$ solvent \\ \hline
      \includegraphics[width=40mm]{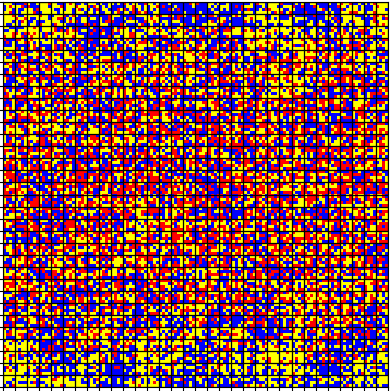} & \includegraphics[width=40mm]{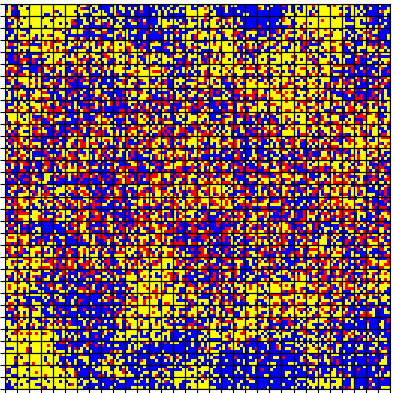} &
      \includegraphics[width=40mm]{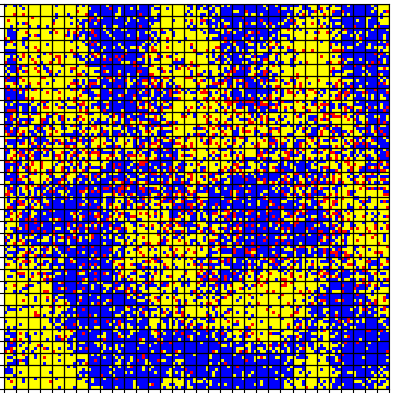} & \includegraphics[width=40mm]{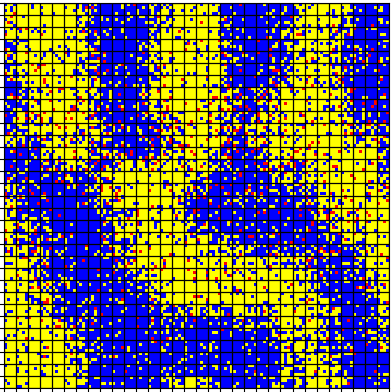} \\
      \small $\phi = 0.6,\ \beta = 0.1$ & $\phi = 0.6,\ \beta = 0.1$ & $\phi = 0.6,\ \beta = 0.1$ &
      $\phi = 0.6,\ \beta = 0.1$ \\
      \\
      \includegraphics[width=40mm]{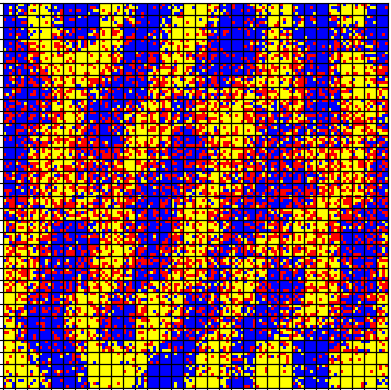} & \includegraphics[width=40mm]{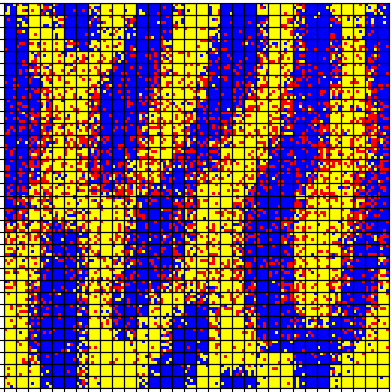} &
      \includegraphics[width=40mm]{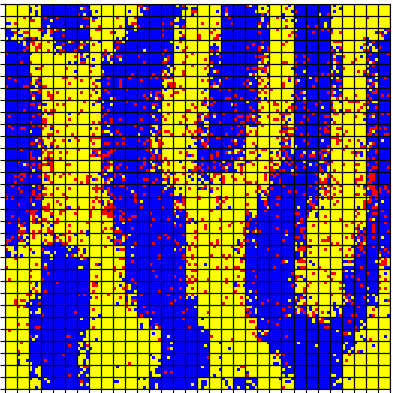} & \includegraphics[width=40mm]{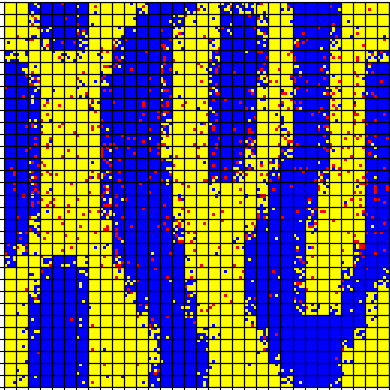} \\
      \small $\phi = 0.6,\ \beta = 0.2$ & $\phi = 0.6,\ \beta = 0.2$ & $\phi = 0.6,\ \beta = 0.2$ &
      $\phi = 0.6,\ \beta = 0.2$ \\
      \\
      \includegraphics[width=40mm]{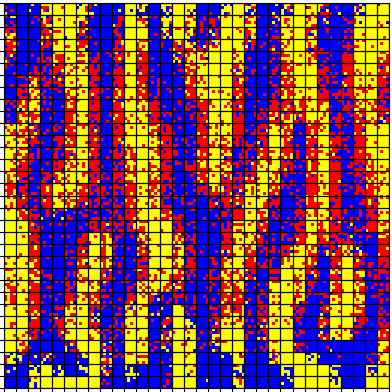} & \includegraphics[width=40mm]{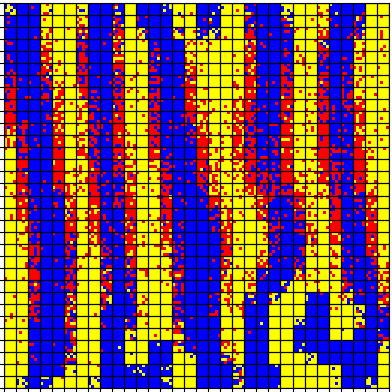} &
      \includegraphics[width=40mm]{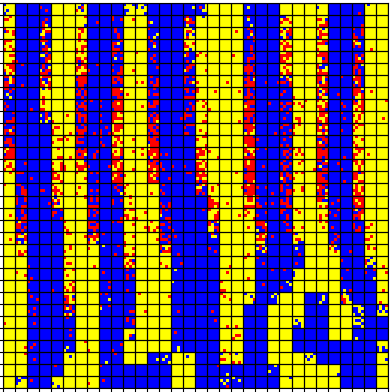} & \includegraphics[width=40mm]{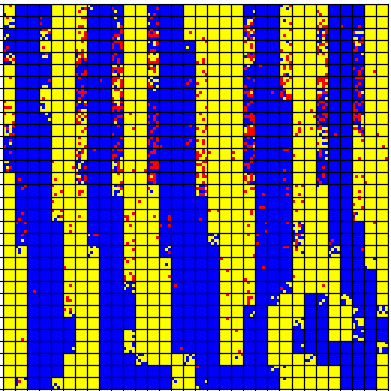} \\
      \small $\phi = 0.6,\ \beta = 0.6$ & $\phi = 0.6,\ \beta = 0.6$ & $\phi = 0.6,\ \beta = 0.6$ &
      $\phi = 0.6,\ \beta = 0.6$ \\ 
\end{tabular}
\caption{First set of simulations. In this case, the volatility parameter is fixed as $\phi = 0.6$ to study the effect of the temperature. In the columns we have 75\%, 50\%, 25\% and 10\% of remaining solvent from the initial amount. We notice that for   high temperature (e.g., $\beta = 0.1$) we do not reach morphology formation. The required number of steps to reach the 10\% of remaining solvent is $8.58 \cdot 10^6$ for $\beta = 0.1$, $8.30 \cdot 10^6$ for $\beta = 0.2$, and $8.51 \cdot 10^6$ for $\beta = 0.6$.}
\label{tab:beta}
\end{figure}

For the first round of simulations, we fix the volatility parameter to $\phi = 0.6$ and look for morphology formations as $\beta$ changes. The other parameters are fixed as above. In each row of Figure \ref{tab:beta}, we display snapshots of the evolution of our systems for different temperatures when the percentage of remaining solvent is the $75\%$, $50\%$, $25\%$ and the $10\%$ of the initial content of solvent.
In this case, with a lattice containing $128 \times 128$ particles of which $40\%$ is solvent, those percentages corresponds to $30\%$, $20\%$, $10\%$ and $4\%$ of the total amount of particles, respectively.
We notice that the temperature plays an important role in the phase separation of the components: if the temperature is extremely high, i.e. in the limiting case in which $\beta$ vanishes, we do not reach any morphology formation. Indeed, in this particular case the thermal energy $k_B T=\beta^{-1}$ is much larger than the spin-spin interaction term $J_{\sigma_\mathbf{x} \sigma_\mathbf{y}}$, in the Hamiltonian (\ref{ham_lambda}), hence spins randomly exchange in an uncorrelated fashion.


As shown in Figure \ref{tab:beta}, we notice the first hints of phase separation if we consider a lower temperature, i.e. a higher value of $\beta$. Already for $\beta = 0.1$ we see a small decrease of the energy in the system during the evaporation process: particles are arranged in interpenetrated formations, but all the species are still mixed in those stains and sharp interfaces between the regions are not very clear/smooth. 
Considering $\beta = 0.2$, we notice a clearer distinction in the phase separation between blue and yellow particles. However, solvent particles are  mixed inside the produced morphologies and it is quite likely that they will move away after a sufficiently long time has elapsed. 
Finally, if we consider a much lower temperature $\beta = 0.6$, the energy finds easier the desired minima. This results in the tendency of the solvent (red) to play the role of perimeter for the blue and yellow colored areas. Since the number of steps is fluctuating, we are unable to notice a clear trend in the number of steps needed to reach the stop parameter. In this case,  we need a range from $8.58 \cdot 10^6$ (for $\beta = 0.1$) to $8.30 \cdot 10^6$ (for $\beta = 0.8$ and $\beta = 0.2$) in the number of iterations to reach the $10\%$ of remaining solvent. The formed lanes shown in Figure \ref{tab:beta} (last row) can be perceived as steady state, since their thickness does not change anymore. 
\begin{figure}[h!]
    \centering
    \includegraphics[width=.5\textwidth]{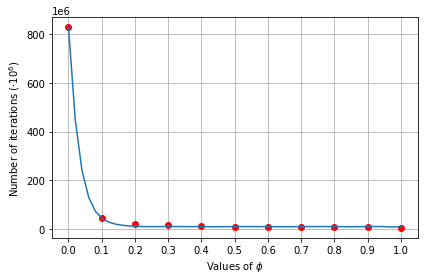}
    \caption{Monotonicity in the number of iterations {\em versus} the volatility parameter $\phi$. This behavior correlates with the fact that simulations are faster when morphologies arrange in vertical strips. The number of iterations goes from $827.24 \cdot 10^6$ (case $\phi = 0$), to $4.96 \cdot 10^6$ (case $\phi = 1$).
    }
    \label{fig:time_phi}
\end{figure}

Having in mind the meaning of the volatility parameter $\phi$, we can speed up the evaporation process by increasing the volatility of the solvent. We see   in Figure \ref{fig:time_phi} the following effect: as we increase $\phi$ for a fixed $\beta$, we notice a strict monotonically decreasing trend in the required number of iterations to reach the stop parameter of 10\% solvent.  The effect is huge already for a small $\phi>0$. The number of iterations goes from $827.24 \cdot 10^6$ for $\phi = 0$, to $4.96 \cdot 10^6$ for $\phi = 1$. We could use a polynomial of order $10$ to interpolate the number of iterations but we notice that we have a better fit with the exponential function $8.17 \cdot 10^8 \cdot e^{-31.46 \cdot x} + 9.70 \cdot 10^6 $ (shown in Figure \ref{fig:time_phi}).
\begin{figure}[h!]
\centering
\noindent
\begin{tabular}{c c c c}
      $75\%$ solvent & $50\%$ solvent & $25\%$ solvent & $10\%$ solvent \\ \hline
      \includegraphics[width=40mm]{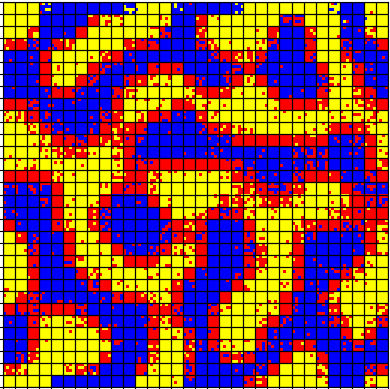} & \includegraphics[width=40mm]{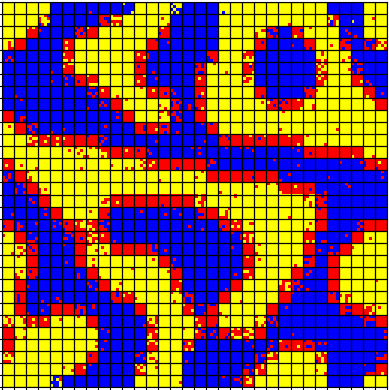} & \includegraphics[width=40mm]{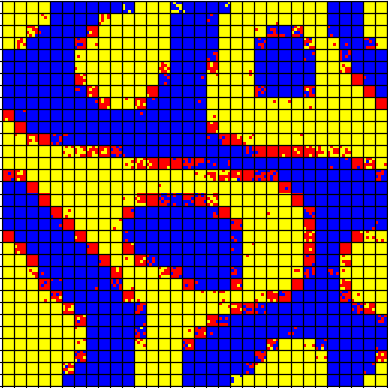} &
      \includegraphics[width=40mm]{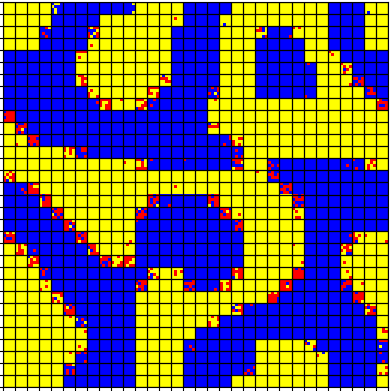} \\
      \small $\phi = 0,\ \beta = 0.8$ & $\phi = 0,\ \beta = 0.8$ & $\phi = 0,\ \beta = 0.8$ &
      $\phi = 0,\ \beta = 0.8$ \\
      \\
      \includegraphics[width=40mm]{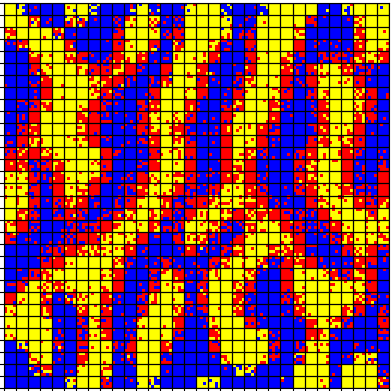} & \includegraphics[width=40mm]{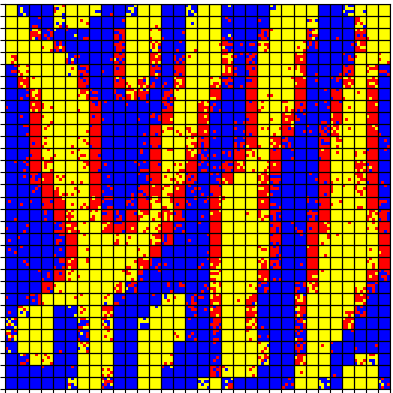} & \includegraphics[width=40mm]{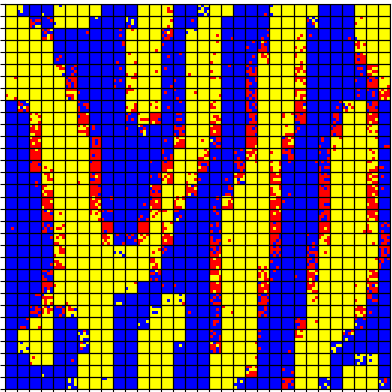} &
      \includegraphics[width=40mm]{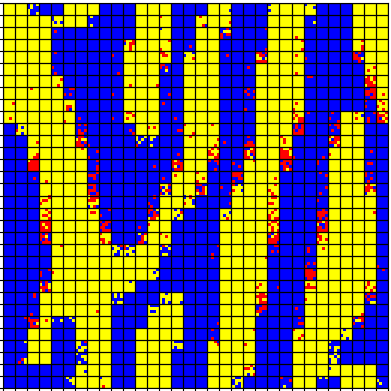} \\
      \small $\phi = 0.1,\ \beta = 0.8$ & $\phi = 0.1,\ \beta = 0.8$ & $\phi = 0.1,\ \beta = 0.8$ &
      $\phi = 0.1,\ \beta = 0.8$ \\
      \\
      \includegraphics[width=40mm]{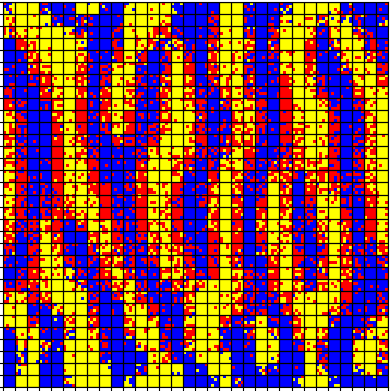} & \includegraphics[width=40mm]{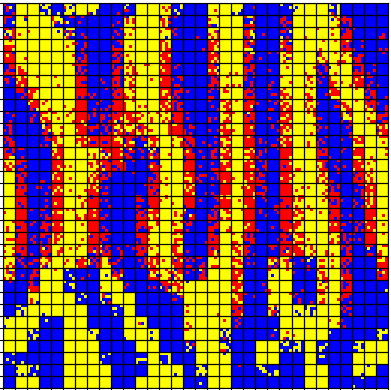} & \includegraphics[width=40mm]{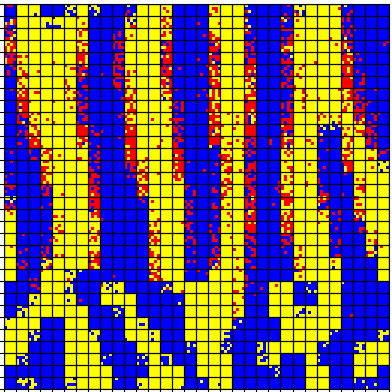} &
      \includegraphics[width=40mm]{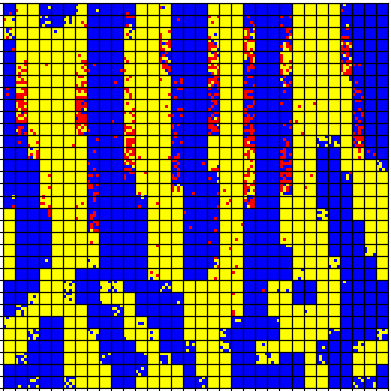} \\
      \small $\phi = 0.5,\ \beta = 0.8$ & $\phi = 0.5,\ \beta = 0.8$ & $\phi = 0.5,\ \beta = 0.8$ &
      $\phi = 0.5,\ \beta = 0.8$ \\
      \\
      \includegraphics[width=40mm]{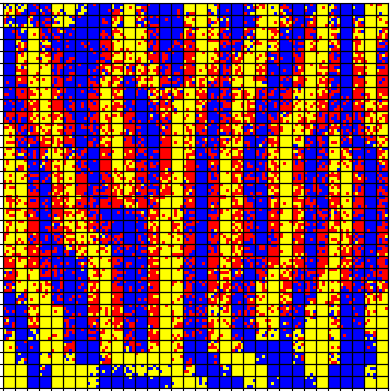} & \includegraphics[width=40mm]{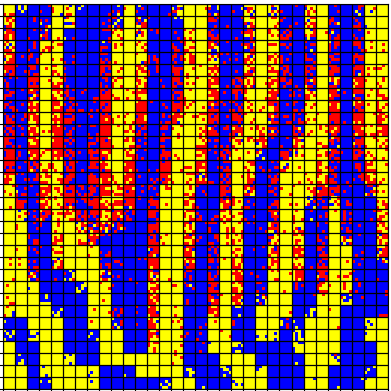} & \includegraphics[width=40mm]{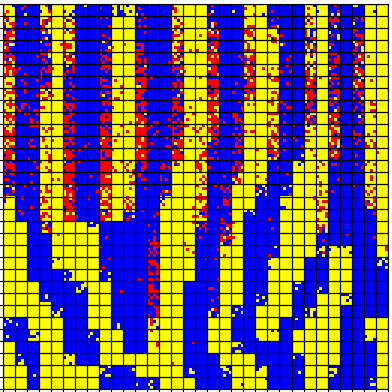} &
      \includegraphics[width=40mm]{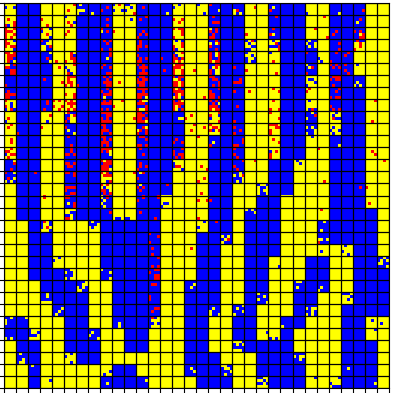} \\
      \small $\phi = 1,\ \beta = 0.8$ & $\phi = 1,\ \beta = 0.8$ & $\phi = 1,\ \beta = 0.8$ &
      $\phi = 1,\ \beta = 0.8$ \\
\end{tabular}
\caption{The temperature is fixed such that $\beta = 0.8$. In the columns we have $75\%$, $50\%$, $25\%$, and $10\%$ of remaining solvent from the initial amount. We notice that for a small value of the volatility parameter ($\phi = 0$) we get a stain-shaped morphologies, while increasing $\phi$ (already from $\phi = 0.1$) we get vertical stripes.}
\label{tab:phi}
\end{figure}
This effect is related to the volatility step of the algorithm: if we consider a high value of volatility parameter, then the probability of the solvent to reach the top of the lattice is higher, otherwise it mostly depends on the Metropolis step, hence on the Hamiltonian (\ref{ham_lambda}). The role of the volatility parameter is not just to speed up (or slow down) the evaporation time. It also plays an important role for the final shape of morphology formations. In Figure \ref{tab:phi}, we study the different morphology shapes obtained for different values of $\phi$ if we fix $\beta = 0.8$. Also in this case we show (from left to right) the evolution of the system when we reach the $75\%$, $50\%$, $25\%$ and $10\%$ of remaining solvent, respectively. In the first row we consider $\phi = 0$. The complete absence of volatility yields the formation of a bi-continuous morphology with no preferential orientation, since the solvent is not forced to go upwards, but only moves upwards in a diffusive mode. The only way for the solvent to reach the top of the lattice depends on $\lambda$, $\beta$ and the Hamiltonian (\ref{ham_lambda}) in the Metropolis step. This also leads to a longer evaporation time, as shown in Figure \ref{fig:time_phi}.
If we choose $\phi = 0.1$, we already see the effect of the volatility, resulting in almost vertical stripes for the morphologies. While for the $75\%$ of remaining solvent we still have some stains, with the evaporation process those stains are deformed in the direction of evaporation.
If we increase the volatility to $\phi = 0.5$, we notice that the vertical stripes are thinner. 
A similar situation arises for the case $\phi = 0$: in the absence of solvent, the two remaining species are not forced anymore to follow the vertical evaporation and they are reorganized following the Metropolis step.
The same effect is better emphasized in the lower half of the lattice, for the case $\phi = 1$, due to a faster evaporation of the solvent. In this limiting case, the vertical stripes are thinner, hence we obtain morphologies with predominant stripe-shaped patterns\footnote{Such vertical morphologies are likely to correspond to a locally-periodic of disc-like patches seen experimentally as top view of the film.}. 
\begin{figure}[b!]
\centering
\noindent
\begin{tabular}{c c c c}
      $75\%$ solvent & $50\%$ solvent & $25\%$ solvent & $10\%$ solvent \\ \hline
      \includegraphics[width=40mm]{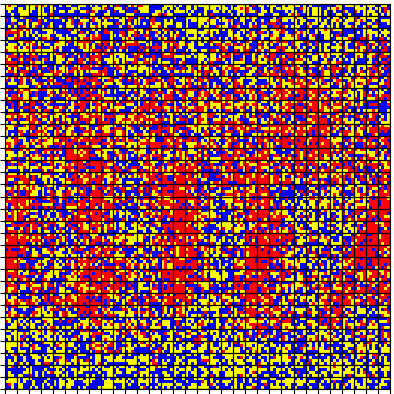} &  \includegraphics[width=40mm]{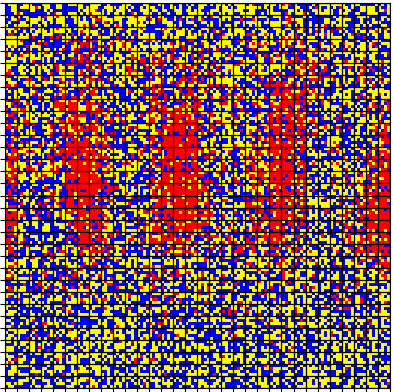} & \includegraphics[width=40mm]{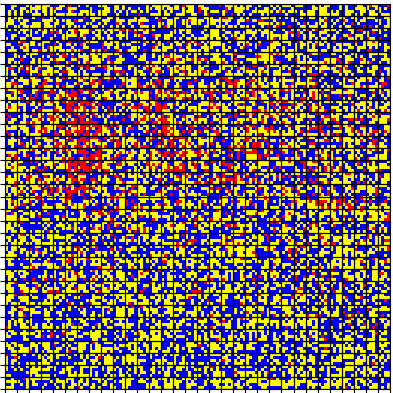} & \includegraphics[width=40mm]{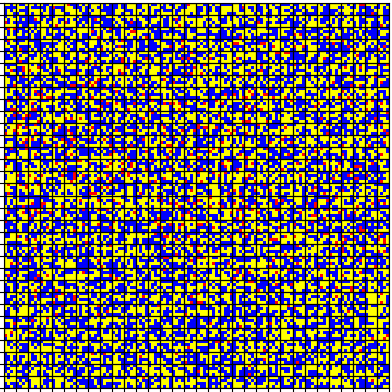} \\
      $J_{+1,-1} = 0.1$ & $J_{+1,-1} = 0.1$ & $J_{+1,-1} = 0.1$ & $J_{+1,-1} = 0.1$ \\
      \\
      \includegraphics[width=40mm]{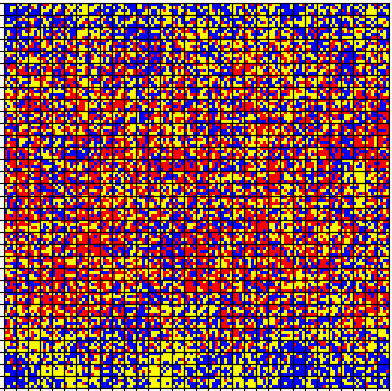} &  \includegraphics[width=40mm]{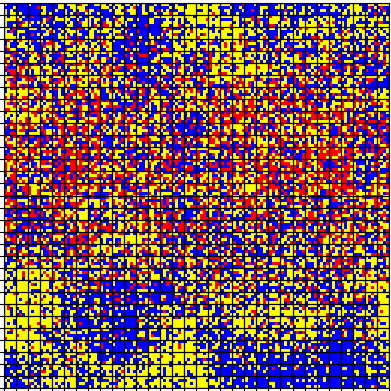} & \includegraphics[width=40mm]{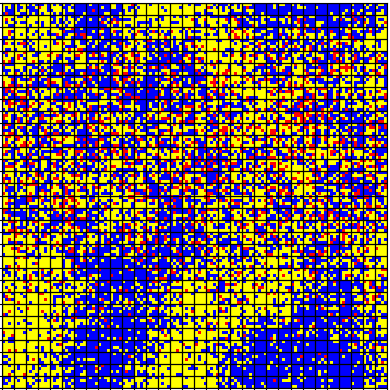} & \includegraphics[width=40mm]{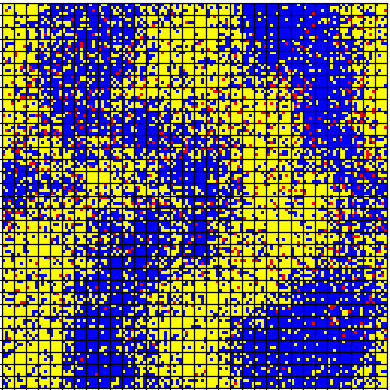} \\
      $J_{+1,-1} = 0.9$ & $J_{+1,-1} = 0.9$ & $J_{+1,-1} = 0.9$ & $J_{+1,-1} = 0.9$ \\
      \\
      \includegraphics[width=40mm]{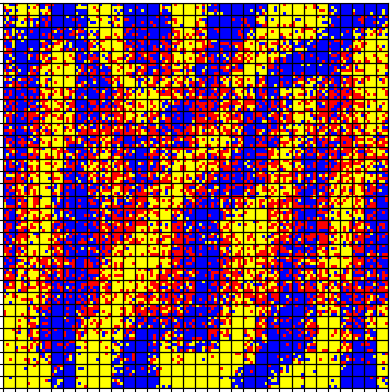} &  \includegraphics[width=40mm]{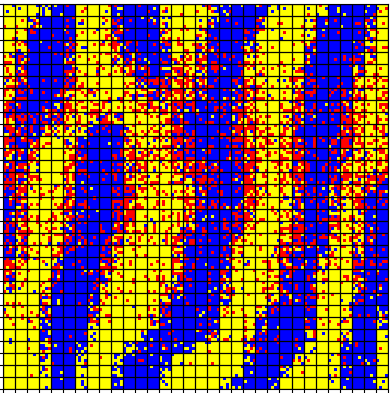} & \includegraphics[width=40mm]{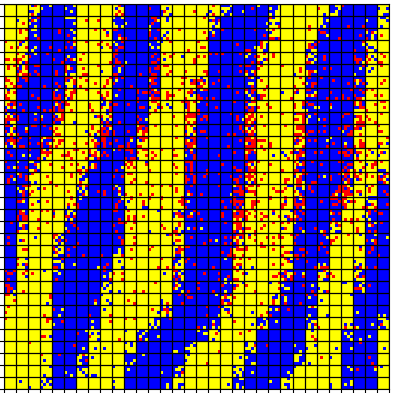} & \includegraphics[width=40mm]{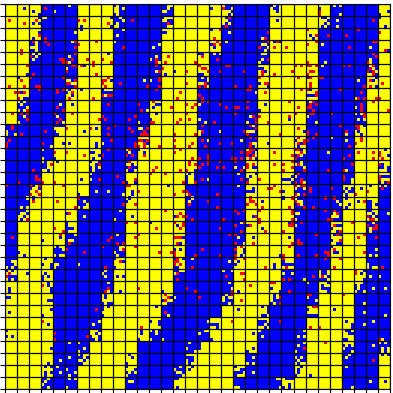} \\
      $J_{+1,-1} = 2$ & $J_{+1,-1} = 2$ & $J_{+1,-1} = 2$ & $J_{+1,-1} = 2$ \\
      \\
      \includegraphics[width=40mm]{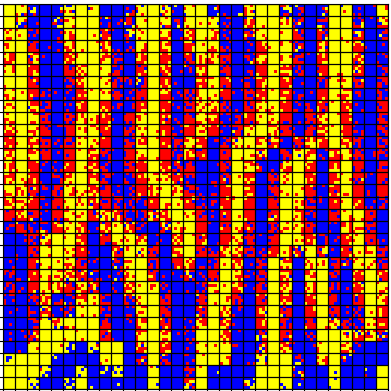} &  \includegraphics[width=40mm]{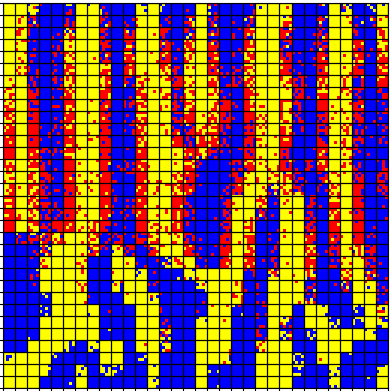} & \includegraphics[width=40mm]{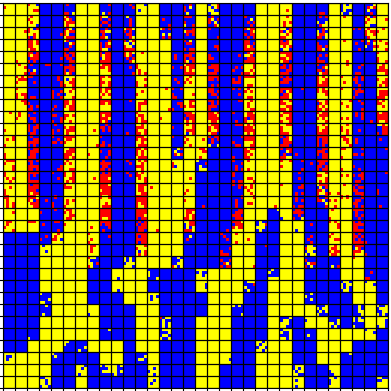} & \includegraphics[width=40mm]{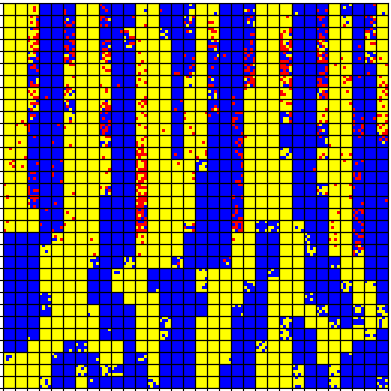} \\
      $J_{+1,-1} = 15$ & $J_{+1,-1} = 15$ & $J_{+1,-1} = 15$ & $J_{+1,-1} = 15$
\end{tabular}
\caption{Fixing temperature and volatility parameters as $\beta = 0.6$ and $\phi = 0.6$,    we change the interaction parameter acting between blue and yellow particles. The columns indicate results for  $75\%$, $50\%$, $25\%$, and $10\%$ of remaining solvent from the initial amount. In the first row, we use $J_{+1,-1} = 0.1$, in the second $J_{+1,-1} = 0.9$, in the third $J_{+1,-1} = 2$, while in the last one we take $J_{+1,-1} = 15$.}
\label{tab:interaction}
\end{figure}

Now, we want to study which  effects can be obtained when varying the value of the interaction parameter between the non evaporating species, namely when changing $J_{+1,-1}$. In Figure \ref{tab:interaction}, we fix $\beta = 0.6$, $\phi = 0.6$, the interaction tensor as
\begin{equation*}
    J := 
    \begin{bmatrix}
    0 & 1 & J_{+1,-1} \\
    1 & 0 & 1 \\
    J_{+1,-1} & 1 & 0
    \end{bmatrix}
\end{equation*}
and we display the evolution of the system for $75\%$, $50\%$, $25\%$ and the $10\%$ of remaining solvent, respectively.
Similarly to the limit case when $\beta$ is going to zero, we see that if we consider $J_{+1,-1} = 0.1 \ll J_{0,+1} = J_{0,-1} = 1$, then we do not reach the separation of phases. In this case, the interaction between the blue and yellow particles is too weak compared to the interaction with the solvent to be able to lead to any morphology formation. 
Increasing this interaction to $J_{+1,-1} = 0.9$, as shown in the second row of Figure \ref{tab:interaction}, we do not notice a distinct phase separation. The presence of blue particles in each yellow area and of yellow particles in each blue area is not negligible, so we do not reach a clearly shaped morphology.
Looking at the case $J_{+1,-1} = 2$, we already see a strong phase separation, mainly composed of eight blue and yellow vertical stripes. 

One may wonder why the morphologies are slanted in this particular case (see third row), and whether this effect is connected to some suitable combinations of model parameters. This is not the case and the observed effect is not robust with respect to changes in parameters. Such skewed orientations seem to appear due to a twofold reason: the blue and yellow particles are constrained to satisfy periodic boundary conditions and the total phase separation takes place rather fast reaching stationarity. Consequently, local agglomerations of blue/yellow particles at the top or bottom boundaries are likely to nucleate fast growing elongated  morphologies.
\begin{figure}[h!]
\centering
\noindent
\begin{tabular}{c c c c}
      $75\%$ solvent & $50\%$ solvent & $25\%$ solvent & $10\%$ solvent \\ \hline
      \includegraphics[width=40mm]{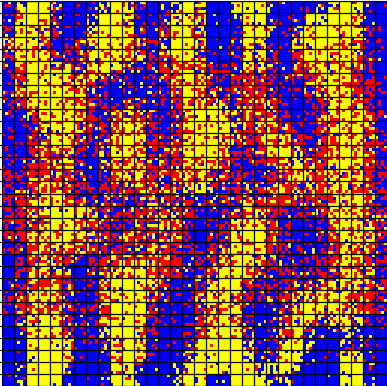} &
      \includegraphics[width=40mm]{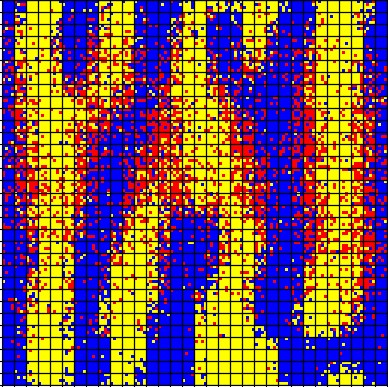} &
      \includegraphics[width=40mm]{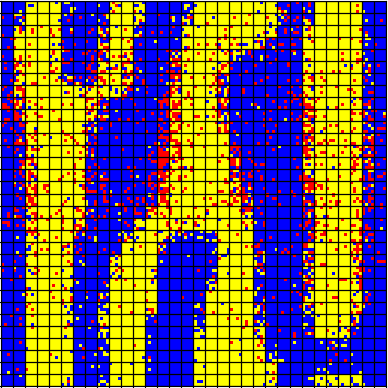} &
      \includegraphics[width=40mm]{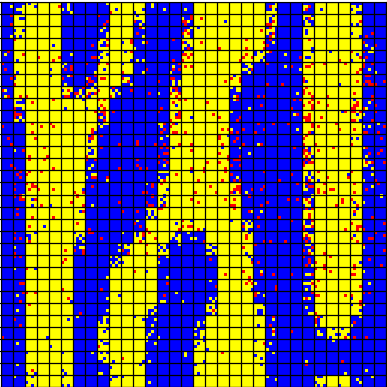} \\
      \footnotesize Periodic & \footnotesize Periodic & \footnotesize Periodic & \footnotesize Periodic\\
      \\
      \includegraphics[width=40mm]{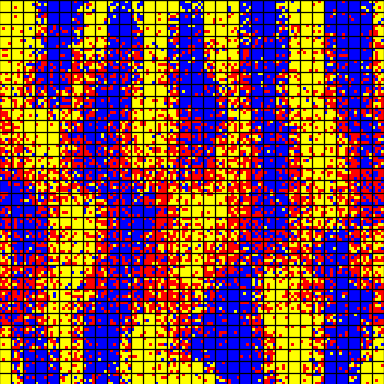} &
      \includegraphics[width=40mm]{basic_simulations/bc/Reflecting_b0.6_phi0.6_L128_J2.0_ratio40.030.030.0_75p.png} &
      \includegraphics[width=40mm]{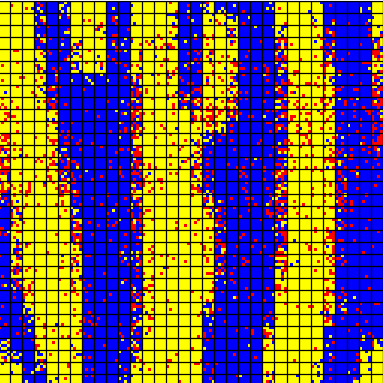} &
      \includegraphics[width=40mm]{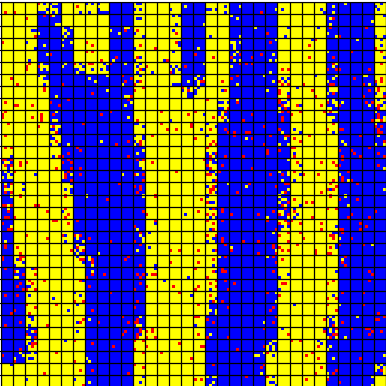} \\
      \footnotesize Reflecting & \footnotesize Reflecting & \footnotesize Reflecting & \footnotesize Reflecting
\end{tabular}
\caption{We chose the case $J_{+1,-1} = 2$. Fixing temperature and volatility parameters as $\beta = 0.6$ and $\phi = 0.6$, we change the boundary conditions. The columns indicate results for  $75\%$, $50\%$, $25\%$, and $10\%$ of remaining solvent from the initial amount. In the first row, we use fully periodic boundary conditions (as defined in Section \ref{model}), while in the second we use reflecting boundary conditions at the top and at the bottom of the lattice.}
\label{tab:refl}
\end{figure}

Apparently, there is something else that plays a stronger role in this skewed formation than periodic boundary conditions: 
the evaporation process has a considerable dependence on initial conditions. 
To clarify this aspect, we show in Figure \ref{tab:refl} the evolution for $75\%$, $50\%$, $25\%$, and $10\%$ of remaining solvent for different boundary conditions on the lattice. The parameters are fixed as in the third row of Figure \ref{tab:interaction}, i.e. fixing $J_{+1,-1} = 2$.
In the first row we present the case with fully periodic boundary conditions, as defined in Section \ref{model}. Here, the parameters and boundary conditions are the same as the third row of Figure \ref{tab:interaction}, however the slightly different initial configuration results in a completely different domain. Somehow, the main direction of the morphologies is vertical even if there is still a skewed branch connecting through the bottom-top periodicity.
The last row of Figure \ref{tab:refl} shows the evolution when bottom-top reflecting boundary conditions are used. We still consider a left-right periodicity, but we do not consider periodic boundary conditions over the top and bottom boundaries, i.e. we do not consider two sites $x,\ y$ in cells $X = (l_1, x_2)$, $Y = (1, x_2)$ as nearest neighbors and we do not compute the energy over the vertical periodicity if the two cells involved in the bond are close to the top or to the bottom. 

It is interesting though to note that, for the chosen parameter regime and regardless of the choice of boundary conditions, the first two panels of each row in Figure  \ref{tab:refl} indicate the occurrence of a dry crust close to the top boundary. As follow-up question we could explore in which way the thickness of such crust depends on key parameters like volatility or temperature.

Considering an even stronger interaction, namely $J_{+1,-1} = 15$, we get a similar effect as for high values of $\phi$: the vertical stripes are thinner in the upper half of the lattice, while in the lower half we notice some horizontal link among the formations, since keeping more areas linked together minimizes the energy of the system.
For bigger values of $J_{+1,-1}$, we see that the vertical stripes are a little bit thinner but the general structure remains the same as the case $J_{+1,-1} = 15$. 

It is worthwhile to also observe the rather thick  lanes of solvent particles pointed out in the first two panels of Figure \ref{tab:interaction} (first row). They appear because the amount of solvent particles in the system is sufficiently high and their interaction strength with respect to the neighboring environment (i.e. the blue and yellow particles) is high. This situation is likely not to happen in the context of organic solar cells. Interestingly though, a very similar situation like the one mentioned here occurs when large crowds of charged colloids \cite{colloids} or of self-driven particles like human crowds \cite{anticipation} or large fish communities \cite{fish} anticipate the position of the wanted exit and collectively adapt their  dynamics to reach it. The top interface, where the solvent evaporates,  plays the role of the exit in such context and the red particles would be the active agents. This type of results indicate that our model is likely to find applications in the context of socio- and econophysics. We will exploit alike connections with population dynamics elsewhere.  

\begin{figure}[b!]
\centering
\noindent
\begin{tabular}{c c c c c}
      $100\%$ solvent & $75\%$ solvent & $50\%$ solvent & $25\%$ solvent & $10\%$ solvent \\ \hline
      \includegraphics[width=30mm]{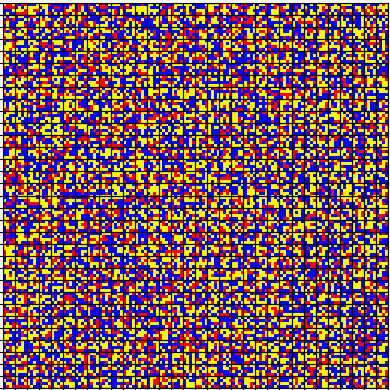} &
      \includegraphics[width=30mm]{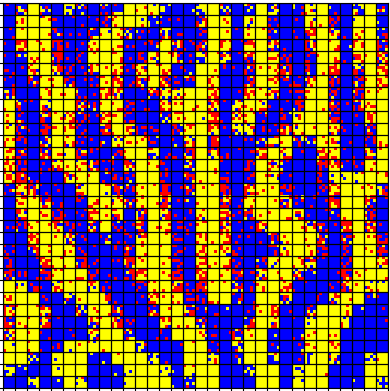} & 
      \includegraphics[width=30mm]{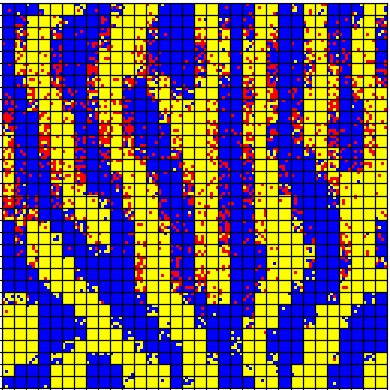} &
      \includegraphics[width=30mm]{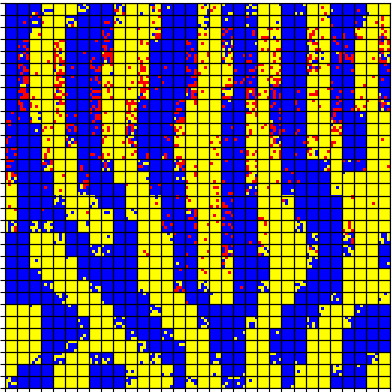} & 
      \includegraphics[width=30mm]{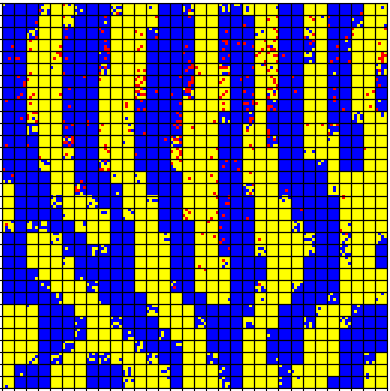} \\
      20:40:40 & 20:40:40 & 20:40:40 & 20:40:40 & 20:40:40 \\ \\
      \includegraphics[width=30mm]{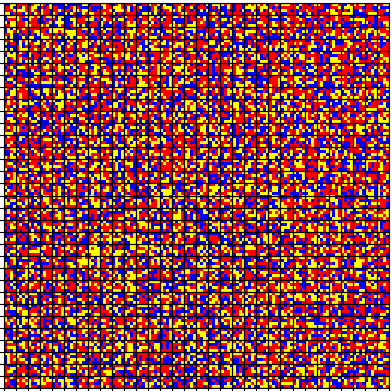} &
      \includegraphics[width=30mm]{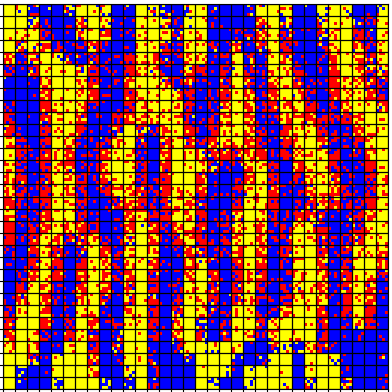} &
      \includegraphics[width=30mm]{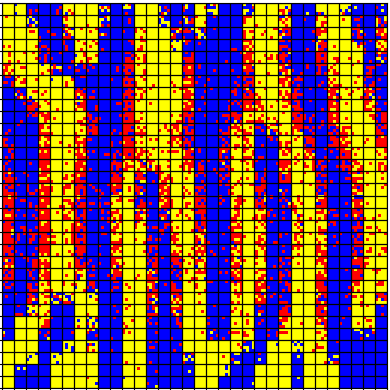} &
      \includegraphics[width=30mm]{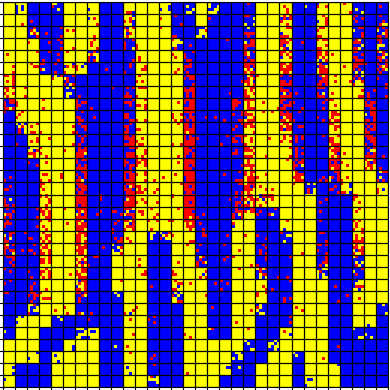} & 
      \includegraphics[width=30mm]{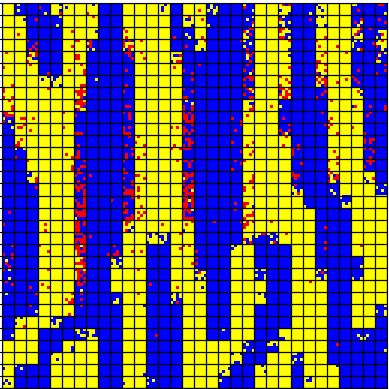} \\
      40:30:30 & 40:30:30 & 40:30:30 & 40:30:30 & 40:30:30  \\ \\
      \includegraphics[width=30mm]{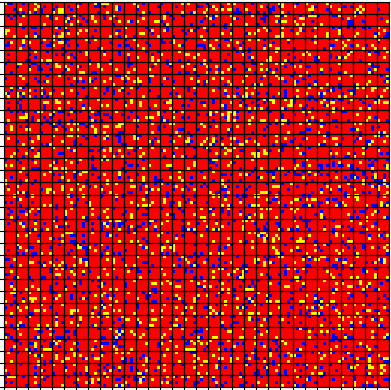} &
      \includegraphics[width=30mm]{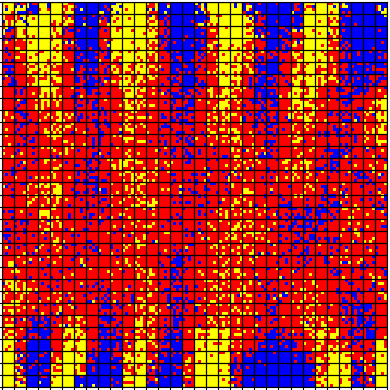} & 
      \includegraphics[width=30mm]{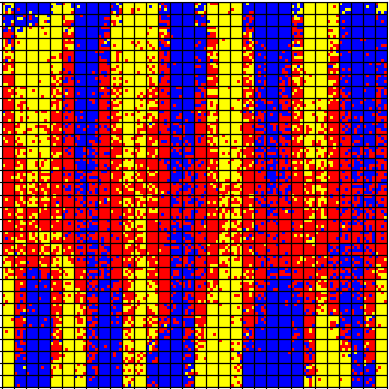} &
      \includegraphics[width=30mm]{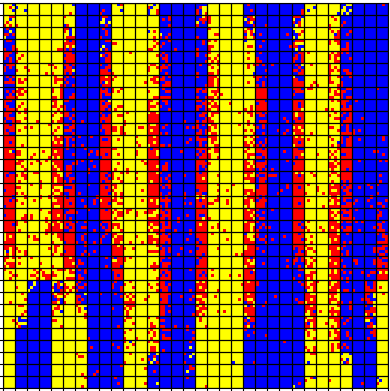} & 
      \includegraphics[width=30mm]{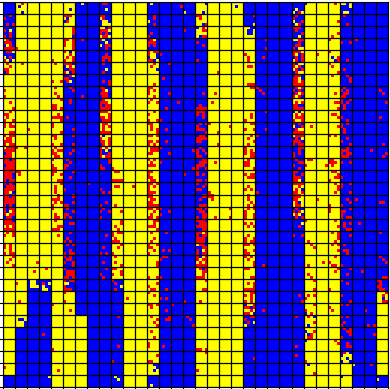} \\
      80:10:10 & 80:10:10 & 80:10:10 & 80:10:10 & 80:10:10
\end{tabular}
\caption{Temperature and volatility parameters are fixed such that $\beta = 0.6$ and $\phi = 0.6$. The interaction tensor is the one defined in (\ref{interaction_simul}). In the columns, we plot the initial configuration (100\% of the initial amount of solvent) and the corresponding evolution with $75\%$, $50\%$, $25\%$, and $10\%$ of remaining solvent from the initial amount. In the first row, we have 20:40:40 as initial ratio of species, for the second 40:30:30, while for the third we take 80:10:10. We notice that for the third case, the vertical stripes are wider due to the considerable migration of solvent.}
\label{tab:ratio}
\end{figure}

As a last simulation in this section, we study how the geometry of the morphologies is affected by choosing different initial ratios of the species involved in the mixture. We fix the temperature and volatility parameters to be $\beta = 0.6$, $\phi = 0.6$, while the interaction tensor is as stated in  (\ref{interaction_simul}). In Figure \ref{tab:ratio},  we show the initial configuration and the corresponding evolution with $75\%$, $50\%$, $25\%$, and $10\%$ of remaining solvent. We recall that the initial ratio is defined by the percentages $p_0 : p_{+1} : p_{-1}$, the considered cases are 20:40:40, 40:30:30 and 80:10:10. We notice that the phase separation is not affected by the initial ratio of species, while the shape of morphologies is altered. However, the evaporation time is higher if we have more solvent in the initial lattice. For a high initial percentage of solvent, we also observe that we have less links among the vertical morphologies.
Starting from the case 20:40:40, we see that the formed morphologies do not follow just the vertical direction of evaporation: in this case the percentage of solvent is too low to force the other two species to align upwards, resulting in a couple of horizontal morphologies. Here, the morphology obtained for the $10\%$ of remaining solvent is quite similar to the one with the $75\%$, due to a fast evaporation process. For this ratio, the $75\%$ of remaining solvent already corresponds to the $15\%$ of total particles.
Already from the case 40:30:30, the stripes are thicker and less linked among them. In this case, the scenario at the $75\%$ of remaining solvent is different than the end, when we have the $10\%$, since the evaporation is much longer than the previous case. Indeed, as specified beforehand, for the case with ratio 20:40:40 the percentage of solvent with respect to the total particles is not enough to drive the formed domain to a considerable different shape.
When we consider 80:10:10 as initial ratio, we observe just eight blue and yellow main stripes in the morphology. This effect is a consequence of the high initial percentage of solvent: even if the evaporation process is longer, the upwards movement of this amount of solvent drives the morphologies to be shaped in a few vertical straight stripes. For this initial ratio, the $75\%$ of remaining solvent is still too much to appreciate a morphology formation, since that corresponds to the $60\%$ of total particles (with respect to the $30\%$ of total particles for the case 40:30:30).

Unlike to what was seen in \cite{Cirillo}, the formed domains look rather similar at top and bottom interfaces due to the boundary conditions.
Moreover, as we notice later on in Section \ref{multiscale}, the width of the formed morphologies is clearly influenced by the size of the chosen mesoscale $\lambda$.

\section{Multiscale effects on morphology formation}\label{multiscale}

In this section, we explore the effects of varying the range of long distance interactions which can be seen at the scale of morphology formation. The parameters that we analyze are the ones involving the size of the lattice (simulation box) and of the cells (spin blocks). We probe different interaction scales for the same choice of reference parameters. In particular, we examine:
\begin{itemize}
    \item repercussions on the scale of the morphologies of the lattice size, by changing $L_1$ and $L_2$;
    \item how the mesoscale length $\lambda$, defining the size of the interaction cells, affects the formation of morphology  for a fixed size of the lattice;
    \item interplay between the parameter $\lambda$ and the lattice size;
    \item effects of the tuning parameter $C$, arising in the Hamiltonian (\ref{ham_lambda}), on both the phase separation and shape of the morphologies. 
\end{itemize}
These numerical experiments are meant to help us understand how to rescale the system so that increasingly larger values of $\lambda$ can be taken without increasing too much the size of the simulation box. The simulations will indicate that there is a clear connection between these two parameters. Similarly as in Section \ref{basic}, the results are explained  by displaying the graphical output of the listed simulations and by noticing the corresponding number of iterations. 

To run the simulations proposed in this section, we fix a target morphology formation, hence we make a selection of the key parameters pointed out in Section \ref{basic}. Particularly, we choose the volatility  $\phi = 0.6$, the temperature $\beta = 0.6$, and the interaction tensor as defined in (\ref{interaction_simul}). For the initial ratio of the mixture components, we set 40:30:30. Moreover, we take as simulation box size $256 \times 256$ and $\lambda = 4$ as cell size. The tuning parameter entering the Hamiltonian (\ref{ham_lambda}) is $C = 1$, unless otherwise specified. In this section, we show the initial configuration in every  presented figure.
\begin{figure}[h!]
\centering
\noindent
\begin{tabular}{c c c}
      $100\%$ solvent & $50\%$ solvent & $25\%$ solvent \\ \hline
      \includegraphics[width=45mm]{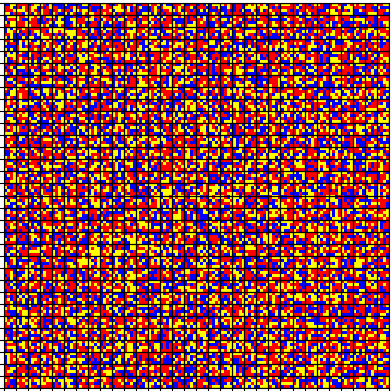} &
      \includegraphics[width=45mm]{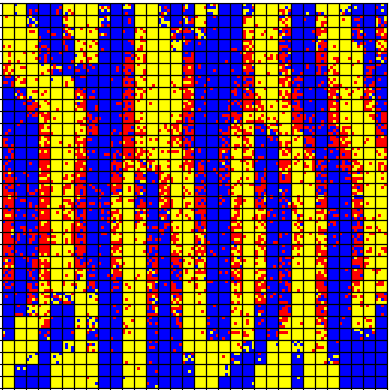} &
      \includegraphics[width=45mm]{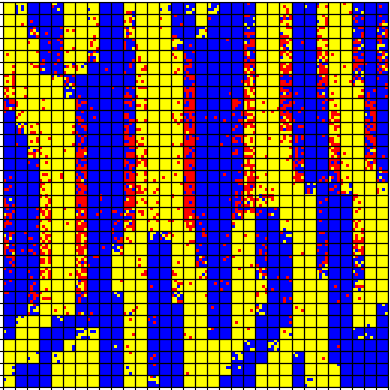} \\
      $128 \times 128$ & $128 \times 128$ & $128 \times 128$ \\ \\
      \includegraphics[width=45mm]{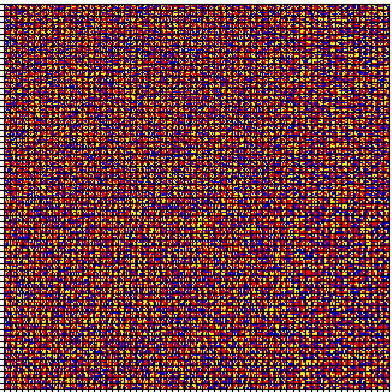} &
      \includegraphics[width=45mm]{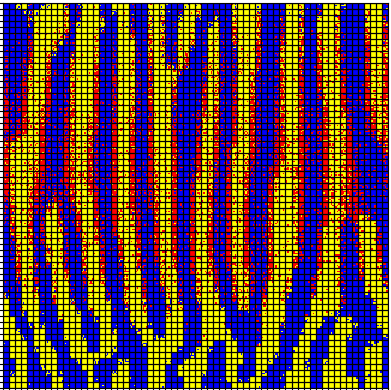} &
      \includegraphics[width=45mm]{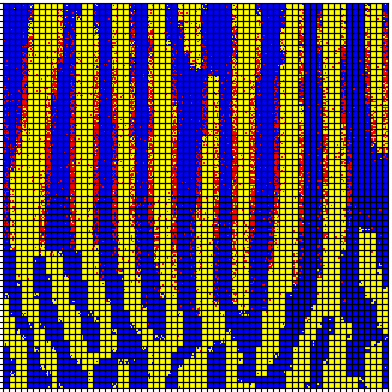} \\
      $256 \times 256$ & $256 \times 256$ & $256 \times 256$ \\ \\
      \includegraphics[width=45mm]{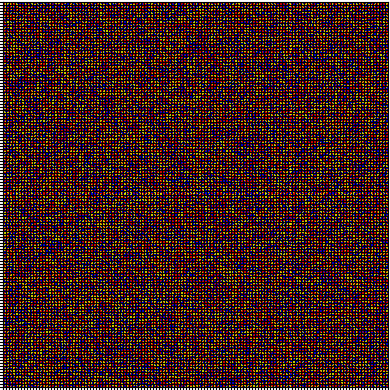} &
      \includegraphics[width=45mm]{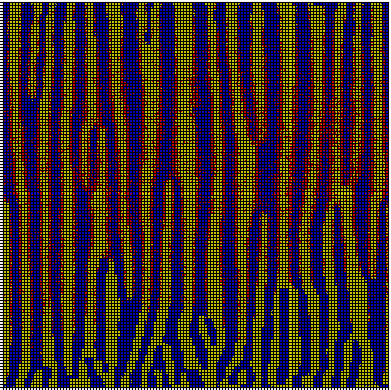} &
      \includegraphics[width=45mm]{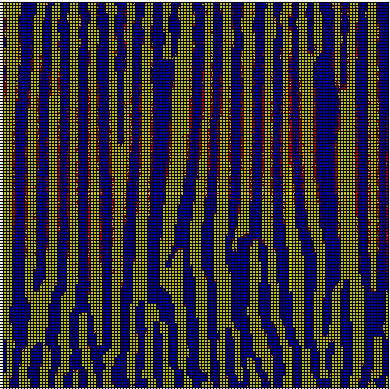} \\
      $512 \times 512$ & $512 \times 512$ & $512 \times 512$
\end{tabular}
\caption{Temperature and volatility parameters are fixed such that $\beta = 0.6$ and $\phi = 0.6$. In the columns we have the initial configuration (100\% of the initial amount of solvent) and the evolution with 50\% and 25\% of remaining solvent from the initial amount. In the first row we consider the size of the lattice to be $128 \times 128$, in the second one $256 \times 256$, in the third one $512 \times 512$.}
\label{tab:size}
\end{figure}

Firstly, we consider the effects of different box sizes. The parameters are fixed as above, except for $L_1$ and $L_2$. In Figure \ref{tab:size}, we display the initial configuration (with 100\% of solvent) and the evolution with 50\% and 25\% of remaining solvent.
In the first row we consider the box size to be $128 \times 128$ (with 6519 red particles), in the second $256 \times 256$ (with 26481 red particles), while $512 \times 512$ (with 104881 red particles) in the third. Clearly, if we consider a lattice with more solvent, the evaporation process will be longer and will exceed a reasonable simulation length; see Section \ref{basic}. Considering that in the case $512 \times 512$ the lattice contains an excessive amount of solvent and it took almost six days on the supercomputer to reach the 25\% of remaining solvent, we avoid for this round of simulations to display the 10\% column.  

It may seem that if we consider a bigger box size we are just doing a ``zoom out'' of our system, but we also notice that the structure of the formed morphologies is slightly different.
We start from the first row of Figure \ref{tab:size}, that displays the usual morphology visible in the last row of Figure \ref{tab:beta} and in the second row of Figure \ref{tab:ratio}.
Moving the attention to the second row, i.e. the case $256 \times 256$, we notice that the average width (in term of lattice sites) of the morphology is the same. However, in the lower half of the lattice we observe more horizontal links among the morphologies than in the upper half. As in the case shown in the last row of Figure \ref{tab:ratio}, the stripes seem to be steeper in the top half of the lattice due to the upwards movement of the great amount of solvent.
Lastly, we see the same behavior if we move from the second to the last row, i.e. the $512 \times 512$ case. Here, the average width of the formed morphologies is still the same as in the first row. Most of the horizontal links among the vertical stripes are concentrated in the lower quarter of the lattice, while in the remaining part of the lattice the shape of morphologies is similar to the upper half of the case $256 \times 256$.
\begin{figure}
    \centering
    \includegraphics[width=.7\textwidth]{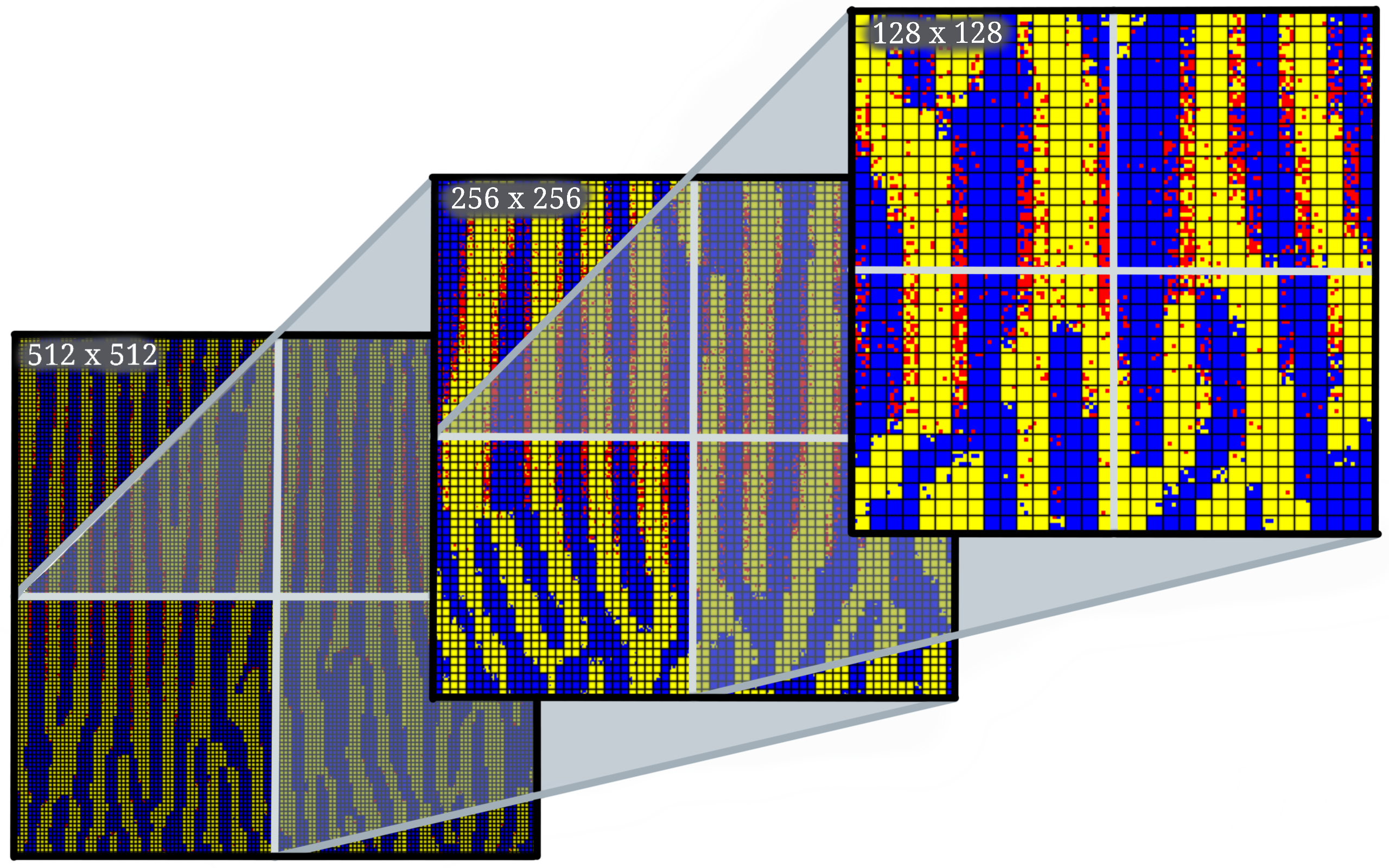}
    \caption{Self-similarity type effect while shrinking the lattice: for a low percentage of remaining solvent the morphology formations in the lower quadrant of a lattice are similar to the one in the lattice with same size of the quadrant; in the upper area the morphologies are mostly vertical stripes.}
    \label{fig:zoom}
\end{figure}
In Figure \ref{fig:zoom} we display this sort of self-similarity type behavior: if we display the $512 \times 512$ lattice with the 25\% of remaining solvent from the last row of Figure \ref{tab:size}, one of its lower quadrants is a $256 \times 256$ lattice with a similar morphology as the last evolution of the $256 \times 256$ configuration in the second row of Figure \ref{tab:size}. We can follow the same reasoning as we shift from one of the lower quadrant of this $256 \times 256$ lattice, that is a $128 \times 128$ lattice, to the proper evolution of the lattice in the first row of Figure \ref{tab:size}.
It is worthwhile to observe that this self-similarity effect holds only if the percentage of remaining solvent is small enough. A similar feature has been pointed out in Figure \ref{tab:size}. 
If we consider the evolution at the 50\% of the studied cases, most of the solvent is still in the central belt of each of the presented lattices. This feature destroys similarity.

\begin{figure}[h!]
\centering
\noindent
\begin{tabular}{c c c c c}
      $100\%$ solvent & $75\%$ solvent & $50\%$ solvent & $25\%$ solvent & $10\%$ solvent \\ \hline
      \includegraphics[width=30mm]{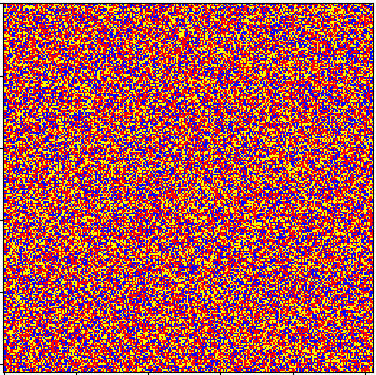} &
      \includegraphics[width=30mm]{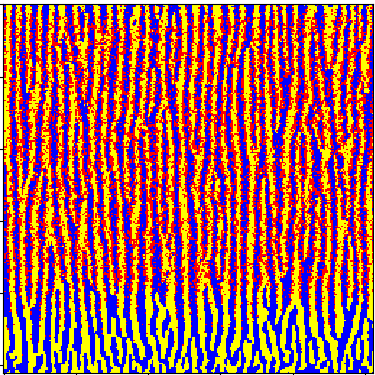} & 
      \includegraphics[width=30mm]{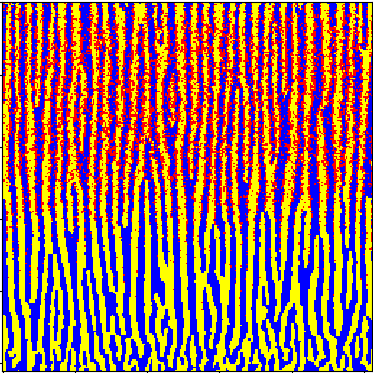} &
      \includegraphics[width=30mm]{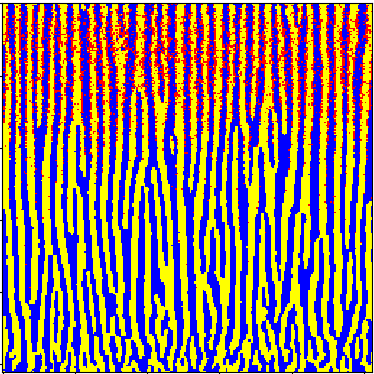} & 
      \includegraphics[width=30mm]{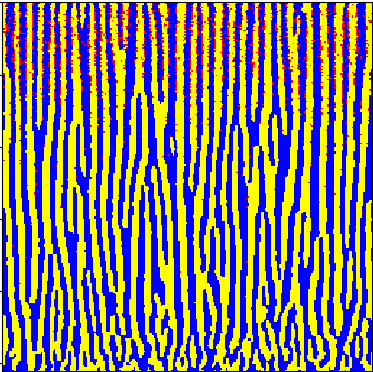} \\
      $\lambda = 1$ & $\lambda = 1$ & $\lambda = 1$ & $\lambda = 1$ & $\lambda = 1$ \\ \\
      \includegraphics[width=30mm]{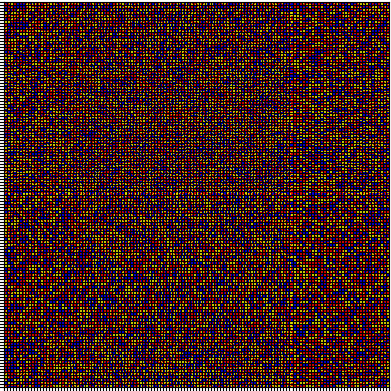} &
      \includegraphics[width=30mm]{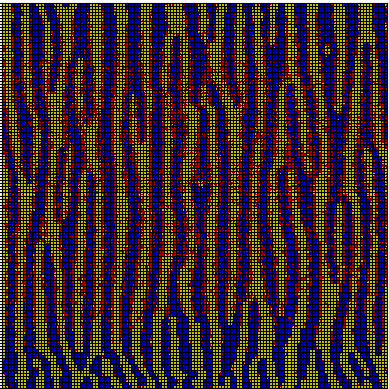} &
      \includegraphics[width=30mm]{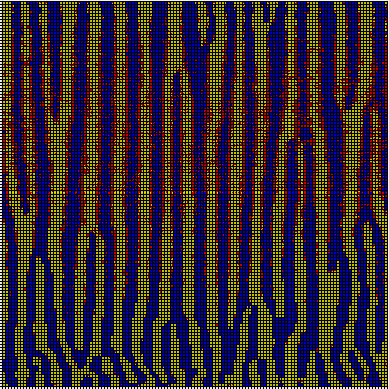} &
      \includegraphics[width=30mm]{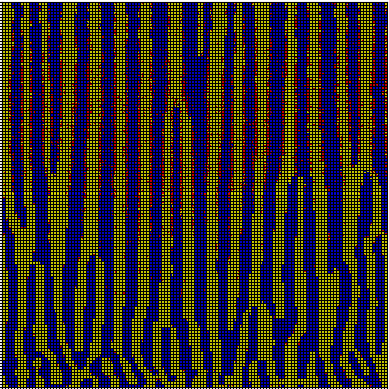} & 
      \includegraphics[width=30mm]{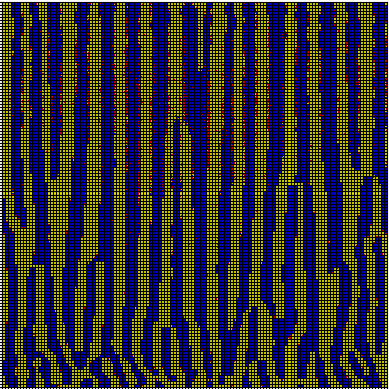} \\
      \small $\lambda = 2$ & $\lambda = 2$ & $\lambda = 2$ & $\lambda = 2$ & $\lambda = 2$ \\ \\
      \includegraphics[width=30mm]{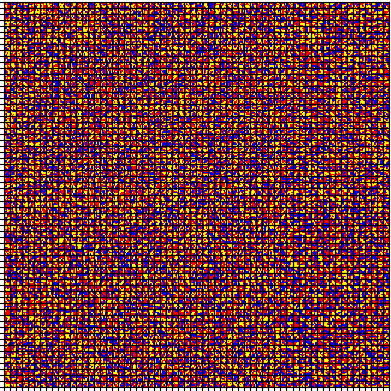} &
      \includegraphics[width=30mm]{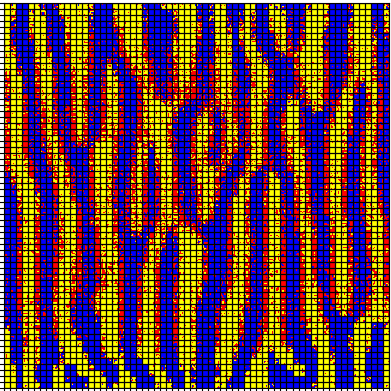} & 
      \includegraphics[width=30mm]{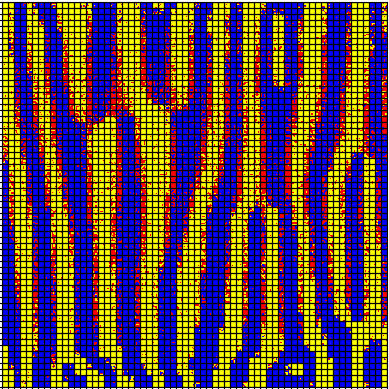} &
      \includegraphics[width=30mm]{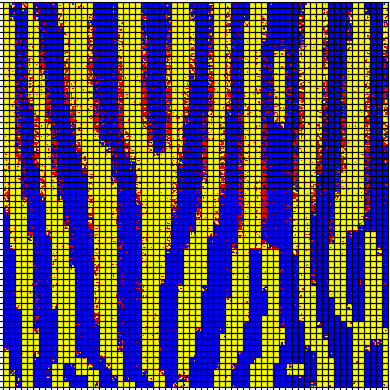} & 
      \includegraphics[width=30mm]{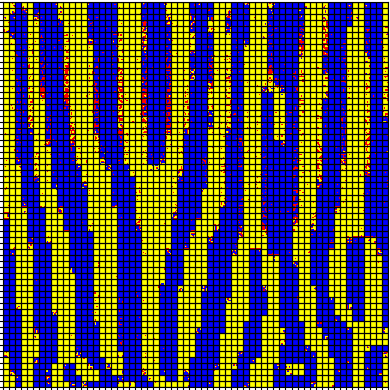} \\
      $\lambda = 4$ & $\lambda = 4$ & $\lambda = 4$ & $\lambda = 4$ & $\lambda = 4$ \\ \\
      \includegraphics[width=30mm]{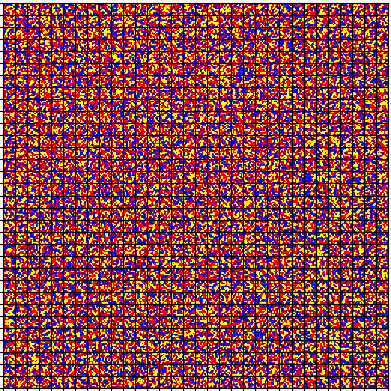} &
      \includegraphics[width=30mm]{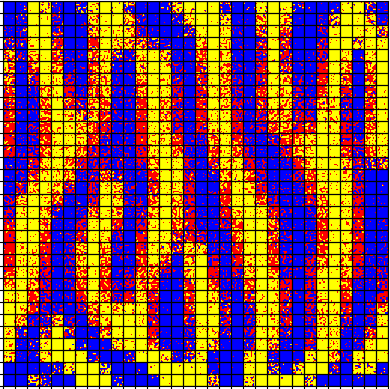} & 
      \includegraphics[width=30mm]{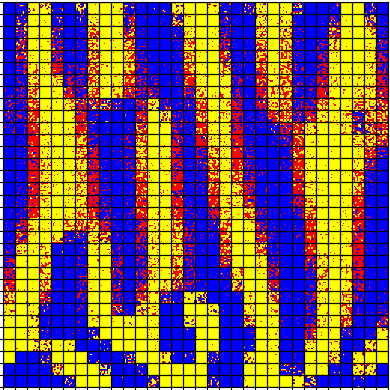} &
      \includegraphics[width=30mm]{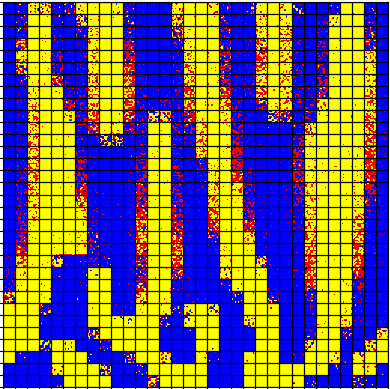} & 
      \includegraphics[width=30mm]{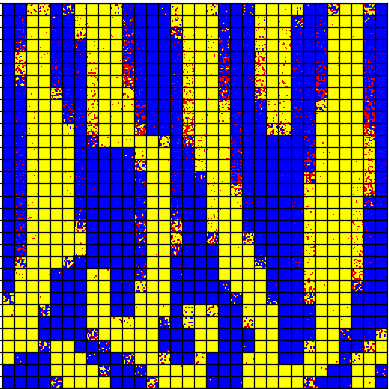} \\
      $\lambda = 8$ & $\lambda = 8$ & $\lambda = 8$ & $\lambda = 8$ & $\lambda = 8$
\end{tabular}
\caption{Parameters are fixed such that $\beta = 0.6$, $\phi = 0.6$, $J$ as in (\ref{interaction_simul}), $C = 1$, and $L_1 = L_2 = 256$. In the columns, we display the initial configuration (100\% of the initial amount of solvent) and the corresponding evolution with $75\%$, $50\%$, $25\%$, and $10\%$ of remaining solvent from the initial amount. In the rows, we set $\lambda = 1$, $\lambda = 2$, $\lambda = 4$, and $\lambda = 8$, respectively.}
\label{tab:lambda}
\end{figure}
We can now switch to considering the effects of different spin block sizes, namely when altering the mesoscopic parameter $\lambda$. We chose the parameters as specified above in the section. Moreover, from the latter paragraph, we know that we can consider a bigger lattice, i.e., with box size $256 \times 256$, with minimal loss of generality. The size choice for this simulation is forced by the values of $\lambda$ that we want to analyze. In  Figure \ref{tab:lambda}, we show the case $\lambda = 2^i,\ i = 0,\ 1,\ 2,\ 3$, while in the column we display the initial configuration and the corresponding evolution with $75\%$, $50\%$, $25\%$, and $10\%$ of remaining solvent. Since we need $\lambda \ll \operatorname{min} \{ L_1,\ L_2 \}$, we prefer to use a bigger box size when it comes to the case $\lambda = 8$. We observe that the size $\lambda$ can also speed up (or slow down) the process: we start with $2.79 \cdot 10^8$ iterations for $\lambda = 1$, then $2.22 \cdot 10^8$ for $\lambda = 2$, subsequently $9.73 \cdot 10^7$ for $\lambda = 4$, and lastly $3.44 \cdot 10^7$ iterations for $\lambda = 8$. For a bigger value of $\lambda$, the evaporation is faster because the particles in the lower area of the lattice can freely move from a box to another, whilst if we consider the case $\lambda = 1$, a particle in the bottom of the lattice has to ``travel'' across every single site before evaporating.
The case $\lambda = 1$, shown in the first row of Figure \ref{tab:lambda}, is widely studied with a microscopic lattice-based model and a short interaction Hamiltonian in \cite{Mario}.
Moving to the second row, we have the case $\lambda = 2$ that presents a thicker morphology than the previous case. We also notice less independent vertical stripes.
In the third row, we can find the case $\lambda = 4$. For this spin block size, the morphologies are even wider and with more horizontal agitation.
Lastly, we have $\lambda = 8$ in the last row. This case seems similar to the cases studied in Section \ref{basic}, as well as the first row of Figure \ref{tab:size}. These similarities come from a wide morphology with similar horizontal links.
As we can understand from the foregoing comments, the spin block size plays an important role in the rescaling of the system, as it defines the lenght scale of interaction in the Hamiltonian (\ref{ham_lambda}). Particularly, increasing this size, we are decreasing the number of stripes in the formations. In fact,  we are increasing the relative thickness of the morphologies with respect to the box size. It is noteworthy that if we use the spin blocks as measurement units, then the average width of the morphologies is the same for the presented values of $\lambda$.

In view of the last two paragraphs, we can spot some interconnections between the rescaling with the box size and the one with the spin block size. If we analyze Figure \ref{tab:size} and Figure \ref{tab:lambda} together, we can conclude that if we consider a $256 \times 256$ lattice and change $\lambda$ accordingly, then we are able to reach morphologies that are similar to those of different box sizes. Specifically, visible correlations can be pointed out between:
\begin{itemize}
    \item $\lambda = 4$ for $L_k = 128,\ k=1,2$ and $\lambda = 8$ for $L_k = 256,\ k=1,2$;
    \item $\lambda = 4$ for $L_k = 512,\ k=1,2$ and $\lambda = 2$ for $L_k = 256,\ k=1,2$.
\end{itemize}

We expect that, while keeping those different proportions between the box and the spin block size, we can rescale our overall system so that the structure of the  morphology shapes is preserved. This would bridge information between two distinct mesoscales.

\begin{figure}[h!]
\centering
\noindent
\begin{tabular}{c c c c c}
      $100\%$ solvent & $75\%$ solvent & $50\%$ solvent & $25\%$ solvent & $10\%$ solvent \\ \hline
      \includegraphics[width=30mm]{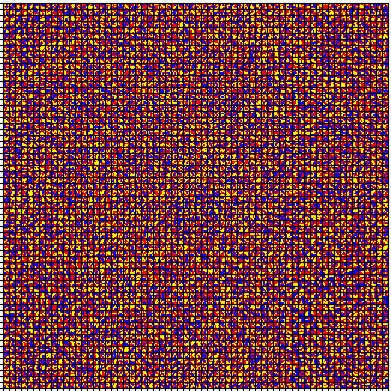} &
      \includegraphics[width=30mm]{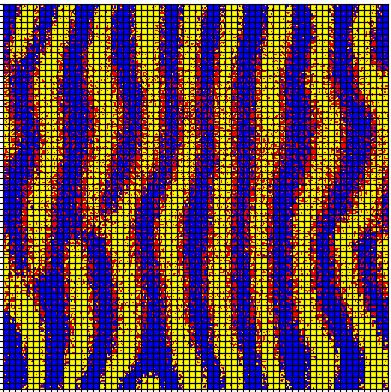} & 
      \includegraphics[width=30mm]{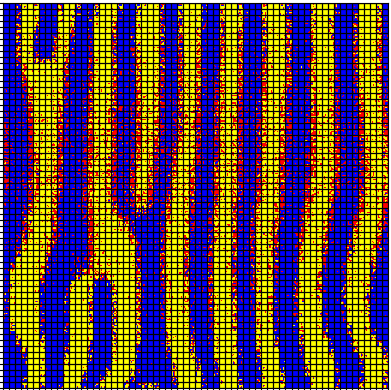} &
      \includegraphics[width=30mm]{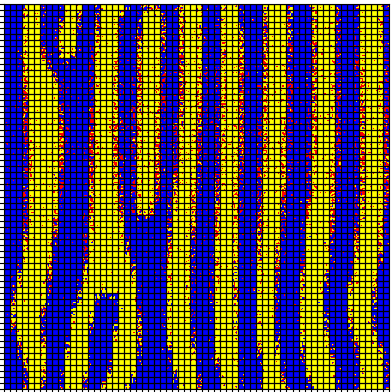} & 
      \includegraphics[width=30mm]{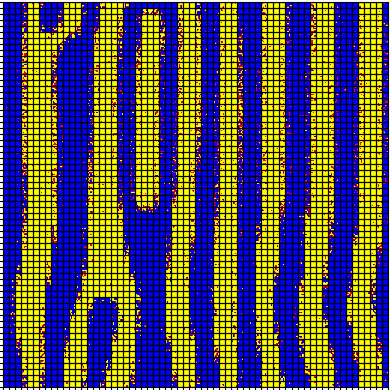} \\
      $C = 0$ &$C = 0$ &$C = 0$ &$C = 0$ &$C = 0$ \\ \\
      \includegraphics[width=30mm]{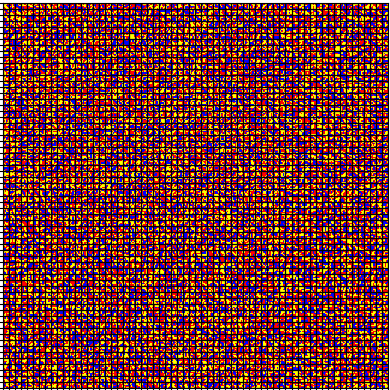} &
      \includegraphics[width=30mm]{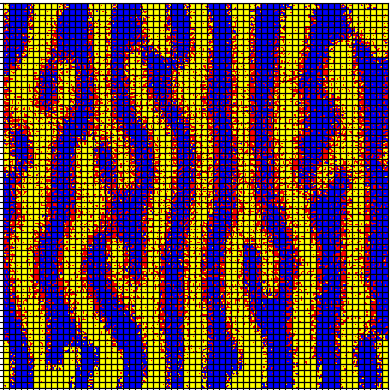} & 
      \includegraphics[width=30mm]{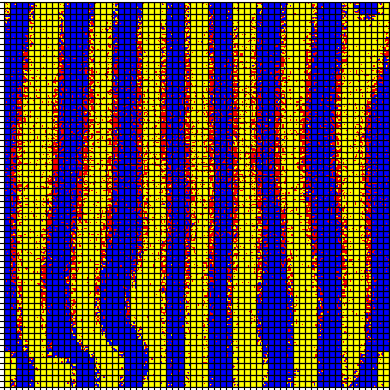} &
      \includegraphics[width=30mm]{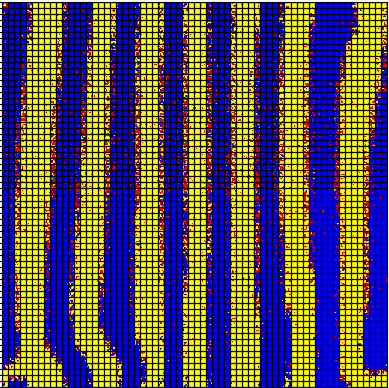} & 
      \includegraphics[width=30mm]{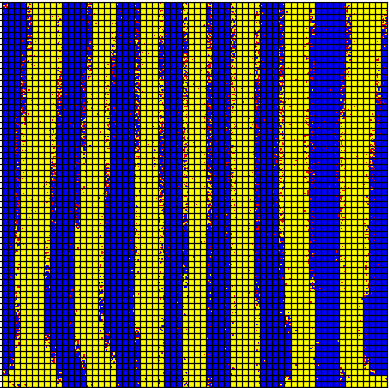} \\
      $C = 0.1$ & $C = 0.1$ & $C = 0.1$ & $C = 0.1$ & $C = 0.1$ \\ \\
      \includegraphics[width=30mm]{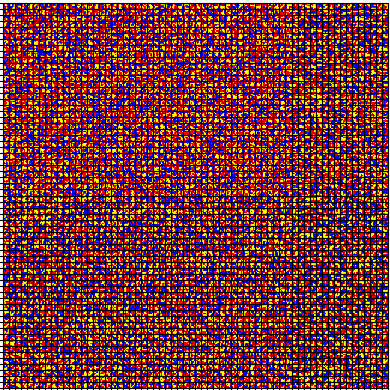} &
      \includegraphics[width=30mm]{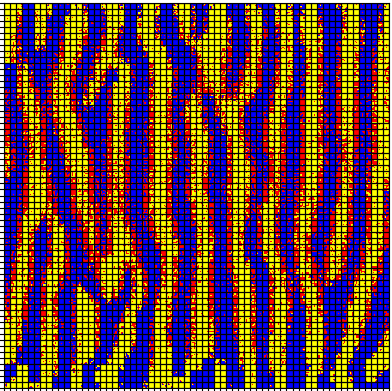} & 
      \includegraphics[width=30mm]{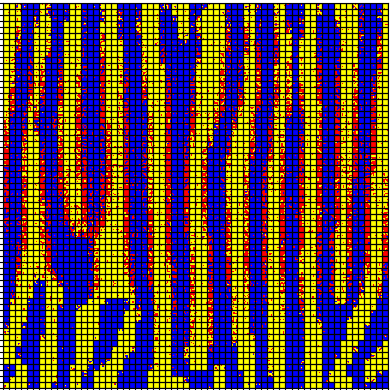} &
      \includegraphics[width=30mm]{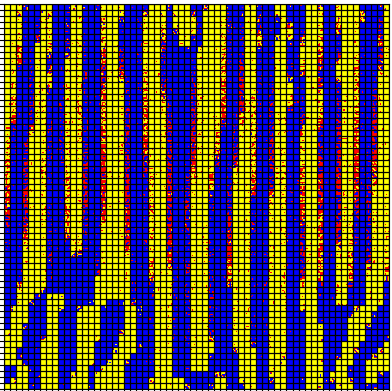} & 
      \includegraphics[width=30mm]{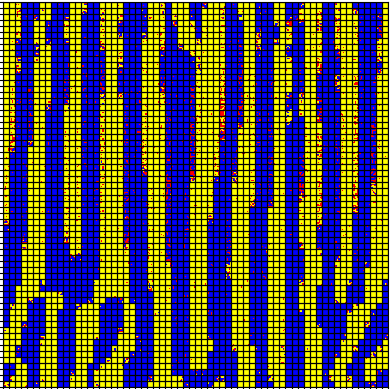} \\
      $C = 1$ & $C = 1$ & $C = 1$ & $C = 1$ & $C = 1$ \\ \\
      \includegraphics[width=30mm]{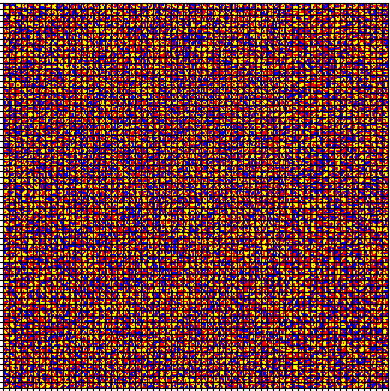} &
      \includegraphics[width=30mm]{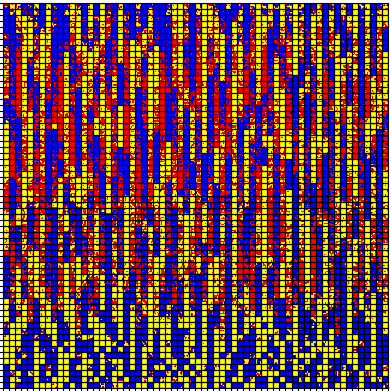} & 
      \includegraphics[width=30mm]{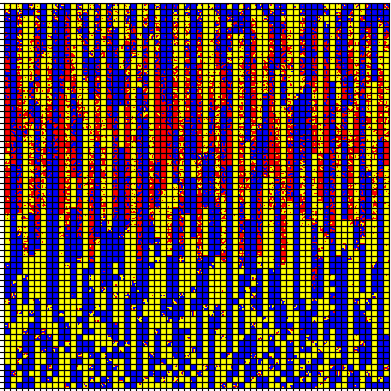} &
      \includegraphics[width=30mm]{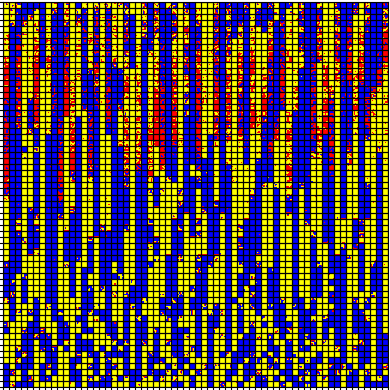} & 
      \includegraphics[width=30mm]{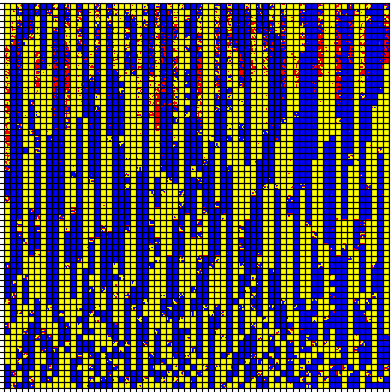} \\
      $C = 10$ & $C = 10$ & $C = 10$ & $C = 10$ & $C = 10$
\end{tabular}
\caption{We fix $\beta = 0.6$, $\phi = 0.6$, $J$ as in (\ref{interaction_simul}), $L_1 = L_2 = 256$, and $\lambda = 4$. In the columns we consider the initial configuration (100\% of the initial amount of solvent) and the corresponding evolution with $75\%$, $50\%$, $25\%$, and $10\%$ of remaining solvent from the initial amount. In the first row we have $C = 0$, in the second $C = 0.1$, in the third $C = 1$, while in the last one $C = 10$.}
\label{tab:C}
\end{figure}
As last simulation, we examine the effect of the tuning parameter $C$, stemming from the Hamiltonian (\ref{ham_lambda}). The purpose of this dimensionless parameter is to weight the intra-cell energy with respect to the total energy. We fix the set of reference parameters as stated in the beginning of this section. We consider the cases $C = 0$, $C = 0.1$, $C = 1$, and $C = 10$. In Figure \ref{tab:C}, we illustrate the initial configuration and the corresponding evolution with $75\%$, $50\%$, $25\%$, and $10\%$ of remaining solvent.
In the first row of Figure \ref{tab:C}, we look at the results for  $C = 0$. Here the morphology formation is striped but, since the intra-cell energy is completely neglected, we can find more solvent than usual in the morphologies, also with some impurities, i.e. blue particles in the yellow zones and vice versa.
In the second row we consider $C = 0.1$.  This case leads to sharper  formations. We also see that now the vertical stripes are really straight.
In the third row, we display the case used for the other simulations when $C = 1$. Using this value for the tuning parameter,  morphologies become visible already from the 75\% of remaining solvent. Finally, the effect of $C = 10$, shown in the last row, is remarkable. Although the phase separation takes place, morphologies seem to meet difficulties to form coherent structures. Most of the spin blocks are filled by particles of the same species, while the location of spin blocks containing the same species is not regular enough to define a morphology. Nonetheless, we did expect this effect to happen as setting $C = 10$ makes the intra-cell energy disproportionate with respect to the inter-cell interaction. If we consider this tuning value in the Metropolis step, hence in the Hamiltonian (\ref{ham_lambda}), the inner energy of every spin block is playing a pivotal role, while the interfacial energy between two cells is almost neglected. We still observe a few vertical links among the spin blocks, but it is mainly because of the upwards movement of the solvent. Even if this situation was foreseeable, it is useful to see the impact on the overall morphology. It would be interesting to unveil a physical interpretation for $C$, of particular importance would be to which extent does $C$ hold information on an eventual $\lambda$-dependence.

\section{Discussion and outlook}\label{outlook}

The simulation tests and the corresponding discussions of the observed effects reported in Section \ref{basic} and Section \ref{multiscale} do not cover all possible scenarios. An exhaustive discussion of the case $\lambda=1$ is done in \cite{Cirillo}. Within the framework of this paper, we aimed: 
\begin{enumerate}
    \item  to show  the capability of the model to produce coherent morphologies at any characteristic mesoscopic length $\lambda$ sufficiently smaller than the simulation box size, and
    \item to explore eventual connections between simulations and morphologies obtained using different values of $\lambda$ and eventually also different volumes of simulation boxes.
\end{enumerate}

Our study opens a number of paths for possible further research exploiting further this multiscale model. We mention here only a few ideas which we deem as being more prominent: 

If one has in mind the applicability of this model to practical questions concerning organic solar cells, then one major difficulty is to set up a computable observable that can be investigated {\em de facto} in an experiment producing morphologies. Usually, as pointed out for instance in \cite{Ellen_AFM_height}, in the laboratory one has  access to a top view height scan of the morphologies as measured by Atomic Force Microscopy (AFM), while our simulations deliver a transversal view on these internal structures. We believe that using a kinetic Monte Carlo approach would allow a better understanding of the physical clock of the morphology formation. Moreover, extending our implementation of our mesoscopic model to a 3D version  would allow a better insight as well as a good match to the AFM images, in which variations in the height of the top surface of the film are shown. For this to happen, besides extending the model, a massive computational effort is needed. Should this be successful, then the triad of methodologies (theoretical, experimental and computational) would then be integrated to make possible an adequate attack of the central two-fold question: Which morphology is best suited for organic solar cells and how to control it?

One of the main aspects that we still wish to investigate further in this context is what type of continuum models are corresponding  to the mesoscopic lattice model formulated here.  Particularly, we would like to explore under which conditions we can bridge suitable averages of our simulation output to what one would obtain with approximating numerically at the continuum macroscopic scale a  Cahn-Hilliard-Cook-type model for a ternary mixture with the evaporation of one component. As first step, we will be investigating which of the geometric structures of the morphologies formed via our mesoscopic simulations can be obtained via changing parameters in the  Cahn-Hilliard-Cook model from \cite{Cook} (or other variants as reported e.g. in \cite{Wodo,Golestanian,Harting}, and, more recently, in \cite{Benoit}), and which are not obtainable via such macroscopic-level simulations. 

Once morphologies obtainable via both the $\lambda$-model as well as by the Cahn-Hilliard-Cook-type system for a ternary mixture with evaporation are classified, then one can think of studying the effect of the obtained morphologies shapes on the efficiency of the macroscopic flux responsible for charge transport. This upscaled information can be reached via averaging the transport of charges over an array of microstructures (REV) which all have as inclusions the obtained morphologies. This discussion can potentially be done at the level of the Nernst-Planck-Poisson system as in \cite{Ray} and eventually it can be combined in a shape optimization framework. A similar work program has been proposed in \cite{Falco}, but they did not consider physically realistic selections of morphologies leaving thus place for a number of improvements.

\section{Acknowledgements} 
MS thanks the {\em InterMaths} graduate program for facilitating his participation in a double degree in applied mathematics between L'Aquila (IT) and Karlstad (SWE). The work of SK (here Vì) took place while they were with Karlstad University. AM thanks SNIC for projects nr. 2020/9-178 and 10-94 (HPC2N) {\em Multiscale simulations of hybrid continuum-discrete-stochastic systems} for providing computational resources and storage capacity as well as VR for project 2018-03648. 
JvS, EM, and SAM acknowledge the funding from the Swedish National Space Agency (Grant 174/19) and the Knut and Alice Wallenbergs Stiftelse (Grant 2016.0059).

\printbibliography 

\end{document}